\documentclass[12pt,a4paper]{article}  
 
\usepackage{epsfig} 
\usepackage{amsmath,amssymb,amsfonts} 
\usepackage{float,a4wide} 
\usepackage{cite} 
\newcommand{\be}{\begin{equation}} 
\newcommand{\en}{\end{equation}}
\newcommand{\bi}{\begin{itemize}} 
\newcommand{\ei}{\end{itemize}}
\newcommand{\bea}{\begin{eqnarray}}
\newcommand{\ena}{\end{eqnarray}}
\newcommand{\Det}{\hbox{Det}}
\newcommand{\hbo}{\hbox to 1 true cm {\hfill } } 
\newcommand{\tr}{\hbox{tr}}

\def\dslash{\partial\kern-.5em\slash}
\def\kslash{k\kern-.5em\slash}
\def\pslash{p\kern-.5em\slash}
\def\Dslash{D\kern-.5em\slash}

\title{Computational Methods in Quantum Field Theory}
\author{Kurt Langfeld \\ \\
School of Mathematics \& Statistics, University of Plymouth \\
Plymouth, PL4 8AA, UK \\
email: kurt.langfeld@plymouth.ac.uk \\
} 
\date{November 19, 2007}

\begin{document} 
\maketitle
 
\begin{abstract}
After a brief introduction to the statistical description of data, 
these lecture notes focus on quantum field theories as they emerge from 
lattice models in the critical limit. For the simulation of these 
lattice models, Markov chain Monte-Carlo methods are widely used. 
We discuss the heat bath and, more modern, cluster algorithms. 
The Ising model is used as a concrete illustration of important 
concepts such as correspondence between a theory of branes and quantum field 
theory or the duality map between strong and weak couplings. 
The notes then discuss the inclusion of gauge symmetries in lattice 
models and, in particular, 
the continuum limit in which quantum Yang-Mills theories arise. 
\end{abstract}

\vfil
\hrule width 5truecm
\vskip .2truecm
\begin{quote}
Notes based on a lecture presented at the XIX Physics Graduate Days 
at the University of Heidelberg, 8th - 12th October 2007. 
\end{quote}

\newpage
\tableofcontents

\section{Statistical data analysis }

\subsection{The central limit theorem }

Assume that we would like to determine a physical observable $\bar{x}$ such 
as a hadron mass or a decay constant by a numerical calculation 
involving statistical methods or by a direct experimental measurement. 
A perfect device would just produce $\bar{x}$ with a single measurement. 
In practice, such a device does not exist. A realistic device produce 
a value $x$ in the interval $[x, x+dx]$ with probability 
$P(x) \, dx$, where the probability distribution $P(x)$ characterises 
the apparatus. We will not assume that our experimental device 
is hampered by systematic errors, but we will assume that the device 
produces the exact value  $\bar{x}$ by an average over many measurements, 
i.e., 
\be 
\int dx \; x \; P(x) \; = \; \bar{x} \; , 
\label{eq:cl1}
\en 
but, depending on $P(x)$, a single measurement can be far from the 
true value. 

\vskip 0.3cm 
As an example, we consider an observable 
$\bar{x}=3$ and a crude experiment which can produce any value for 
between $0$ and $6$ with equal probability: 
\be
P(x) \; = \; \left\{ \begin{array}{ll} 
1/6 & \hbox{ for} \; x\in [0,6] \\
0   & \hbox{ otherwise.} 
\end{array} \right. 
\label{eq:cl1b}
\en
Obviously, a single measurement for $x$ is not sufficient to reveal 
the true observable. The only thing we can do is to repeat the 
measurement $n$ times and to consider the average: 
$$ 
y \; = \; \frac{1}{n} \, \Bigl[ x_1 \; + \; \ldots \; + \; x_n 
\Bigr] \; , 
$$
where $x_1$ to $x_n$ are the values obtained from each of the 
measurements. For the moment, we will assume that the measurements 
are {\it independent}, i.e., that the probability for finding 
a set $\{x_1 \ldots x_n\}$ of data is given by:
$$
P(x_1, \ldots , x_n) \; = \;  P(x_1) \; \ldots \; P(x_n) \; . 
$$
The crucial question is to which accuracy have we 
estimated the true observable $\bar{x}$? 

\vskip 0.3cm 
The answer can be inferred from the probability distribution $Q(y)$ 
for the value $y$:
\be 
Q_n (y) \; = \; \int \prod _{i=1}^n dx_i \; \delta \left( 
y \;  - \; \frac{1}{n} \; [ x_{1} \; + \; \ldots \; + \; x_{n} ]
\right) \; P(x_1) \; \ldots \; P(x_n) \; . 
\label{eq:cl2} 
\en 
Given the proper normalisation of the single event distributions, i.e., 
$$ 
\int dx_i \; P(x_i) \; = \; 1 \; , 
$$
using (\ref{eq:cl1}), 
we can easily show that the average of $y$ coincides with the observable: 
\bea 
\bar{y} &=& \int dy \; y \; Q_n(y) 
\nonumber \\ &=& \int dy \; 
\int \prod _{i=1}^n dx_i \; \frac{1}{n} \; [ x_{1} \; + \; \ldots \; 
+ \; x_{n} ] \; \delta \left( 
y \;  - \; \frac{1}{n} \; [ x_{1} \; + \; \ldots \; + \; x_{n} ]
\right) \; P(x_1) \; \ldots \; P(x_n) \; . 
\nonumber \\ 
&=& \int \prod _{i=1}^n dx_i \; \frac{1}{n} \; [ x_{1} \; + \; \ldots \; 
+ \; x_{n} ] \; \; P(x_1) \; \ldots \; P(x_n) \; = \; 
 \frac{1}{n} \; n \; \int dx \; x \; P(x) \; = \; \bar{x} \; . 
\nonumber 
\ena 
A natural measure for the error $\sigma $ of our estimate is provided by the 
second moment: 
\be 
\sigma ^2 (n) \; = \; \int dy \; \Bigl(y \, - \, \bar{y} \Bigr)^2 \; 
Q_n(y) \; , \hbo \hbox{(the variance)} \; . 
\label{eq:cl3} 
\en 
If the distribution $Q_n(y)$ peaks around the true value for our 
observable $\bar{x}$ and $\sigma (n)$ is tiny, it would mean that 
a single estimator $y$ has high probability to fall close to $\bar{x}$ with 
high probability implying that it yields a good approximation to $\bar{x}$. 

\vskip 0.3cm 
Let us study the moments of the distribution 
$Q(y)$:
\be 
q_m \; = \;  \int dy \; Q_n(y) \; y^m \; . 
\label{eq:cl4} 
\en 
In order to draw further conclusions, we need to restrict the 
classes of single event probability distributions $P(x)$:  
we will assume that its Fourier transform 
\be 
\bar{P}(\beta ) \; = \; 
 \int dx \; P(x) \; \exp \left\{ - i \, \beta \, x 
\right\} 
\label{eq:cl5} 
\en 
is an analytic function of $\beta $ at $\beta =0$. 
As a consequence, the moments of $P(x)$ exist and are given by: 
\be 
p_m \; = \;  \int dx \; P(x) \; x^m \; = \; i^m \; 
\frac{d^m}{d\beta ^m} \bar{P}(\beta ) \; \vert _{\beta = 0 } \; .
\label{eq:cl6} 
\en 
We will further assume that $\bar{P}(\beta )$ vanishes 
for $\vert \beta \vert \to \pm \infty $. 
This seems to be quite a weak constraint. I point out, however, 
that systems with rare but large fluctuations generically fail to possess 
higher moments. One example is stock market indices~\cite{stock_m}. 

\vskip 0.3cm 
Our aim is to express the moments of $Q_n(y)$ in terms of the moments 
of $P(x)$. For this purpose, we rewrite the $\delta$-function in 
(\ref{eq:cl2}) as 
\be
\delta \left( 
y \;  - \; \frac{1}{n} \; [ x_{1} \; + \; \ldots \; + \; x_{n} ]
\right) \; = \; \int \frac{d\alpha }{2\pi } \; \exp [i \, \alpha \, y ] 
\; \prod _{i=1}^n \exp \left\{ - i \, \frac{\alpha }{n} \, x_i 
\right\} \; , 
\label{eq:cl6b} 
\en 
and find 
\bea
Q_n (y) &=&  \int \frac{d\alpha }{2\pi } \; \exp (i \, \alpha \, y ) \; 
\left[ \int dx \; P(x) \; \exp \left\{ - i \, \frac{\alpha }{n} \, x
\right\} \right]^n \; , 
\label{eq:cl7} \\ 
&=& \int \frac{d\alpha }{2\pi } \; \exp (i \, \alpha \, y ) \; 
\left[  \bar{P}\left( \frac{ \alpha }{n} \right) \, 
\right]^n \; . 
\label{eq:cl8} 
\ena
The moments of $Q_n (y)$ are then obtained from 
\be 
q_m \; = \;  \int dy \; \int \frac{d\alpha }{2\pi } \; (-i)^m \; 
\frac{d^m}{d\alpha ^m } \Bigl[ \exp (i \, \alpha \, y ) \, \Bigr] \; 
 \bar{P}^n \left( \frac{ \alpha }{n} \right) \; . 
\label{eq:cl9} 
\en 
After a series of partial integrations with respect to $\alpha $ 
(note that boundary terms vanish by virtue of our above assumptions), 
the latter equation is given by 
\bea 
q_m &=& \int dy \; \int \frac{d\alpha }{2\pi } \; 
\exp (i \, \alpha \, y ) \; (i)^m \; 
\frac{d^m}{d\alpha ^m } \Bigl[   \bar{P}^n \left( \frac{ \alpha }{n} \right) 
\Bigr] 
\nonumber \\ 
&=& \int \frac{d\alpha }{2\pi } \; \frac{1}{n^m} \; \int dy \; 
\exp (i \, \alpha \, y ) \; (i)^m \; 
\frac{d^m}{d\beta ^m } \Bigl[   \bar{P}^n \left( \beta \right) \Bigr]  
\; = \; 
\frac{i^m}{n^m} 
\frac{d^m}{d\beta ^m } \Bigl[   \bar{P}^n \left( \beta \right) \Bigr] 
\; \vert _{\beta =0 } \; .
\label{eq:cl10} 
\ena 
Of particular interest are the so-called cumulants $c_k[Q_n]$ 
of the distribution $Q_n(y)$. These are defined via the generating 
function 
\be 
T_Q(x) \; = \; \sum _{m=0} \frac{1}{m!} \; q_m \; x^m \; , \hbo 
c_k[Q_n] \; := \; \frac{d^k}{dx^k} \; \ln T_Q(x) \; \vert _{x=0} \; . 
\label{eq:cl11} 
\en 
Note that in particular we find for the 'error' $\sigma $ in (\ref{eq:cl3}) 
\be 
\sigma ^2 \; = \; q_2 \; - \; q_1^2 \; = \; c_2[Q_n] \; . 
\label{eq:cl12} 
\en 
Using Taylor's theorem and the explicit expression (\ref{eq:cl10}), we find 
\be 
T_Q(x) \; = \;  \bar{P}^n \left( \frac{ i \, x }{n} \right) \; , \hbo 
c_k [Q_n] \; = \; \frac{i^k}{n^{k-1}} \;  \Bigl[ \ln \bar{P}(0) 
\Bigr]^{(k)} \; , 
\label{eq:cl15} 
\en 
where $(k)$ denotes the $k$th derivative. 
Introducing the cumulants $c_k[P]$ of the single event distribution 
as well, i.e., 
\be 
c_k [P] \; = \; i^k \;  \Bigl[ \ln \bar{P}(0) 
\Bigr]^{(k)} \; ,  
\label{eq:cl15b} 
\en 
we arrive at a very important result: 
\be 
c_k [Q_n] \; = \; \frac{1}{n^{k-1}} \;  c_k [P]\; . 
\label{eq:cl16} 
\en 
Note that the cumulants $ c_k [P] $ are finite numbers which characterise 
the single event probability distribution. Equation (\ref{eq:cl16})
then implies that for increasing number of measurements $n$, 
the higher $(k>1)$ cumulants of $Q_n(y)$ vanish. In particular, 
we find that 
\be 
\sigma (n) \; = \; \sqrt{ c_2 [Q_n] } \; = \; 
\frac{ \sqrt{ c_2 [P] } }{ \sqrt{n} } \; \propto \; 1 / \sqrt{n} \; . 
\label{eq:cl17} 
\en 
For the above example, we find 
\be
p_1 \; = \; \frac{1}{6} \int_0^6 dx \; x \; = \; 3 \; , \hbo 
p_2 \; = \; \frac{1}{6} \int_0^6 dx \; x^2 \; = \; 12 \; , \hbo 
c_2[P] \; = \; 12 - 3^2 \; = \; 3 \; , 
\label{eq:cl17b} 
\en 
and therefore 
$$
\sigma (n) \; = \; \sqrt{3/n} \; . 
$$

\vskip 0.3cm 
It is well known that if $c_k[G]=0$ for $k>2$, the probability distribution 
$G$ is a Gaussian. We therefore expect that if $n$ is chosen 
sufficiently large so 
that we can neglect  $c_k[Q_n]$ with $k >2$, we should be able to approximate 
$Q_n(y)$ by a Gaussian. To support this claim 
(without mathematical rigour), we start from (\ref{eq:cl8}): 
$$
Q_n (y) \; = \; \int \frac{d\alpha }{2\pi } \; \exp (i \, \alpha \, y ) \; 
\exp \left\{ n \; \ln \left[  \bar{P}\left( \frac{ \alpha }{n} \right) \, 
\right] \right\} \; , 
$$
and expand the logarithm with respect to $\alpha $: 
\bea
Q_n (y) &=& \int \frac{d\alpha }{2\pi } \; \exp (i \, \alpha \, y ) \; 
\exp \left\{ n \; \sum _{k=0}^\infty  \frac{1}{k!} 
\left[ \ln \bar{P}(0) \right]^{(k)} \, \left( \frac{\alpha }{n} \right)^k 
\, \right\} 
\nonumber \\ 
&=& 
\int \frac{d\alpha }{2\pi } \; \exp (i \, \alpha \, y ) \; 
\exp \left\{ n \; \sum _{k=1}^\infty \, \frac{1}{k!} \; c_k[P] \; 
\left( -i \, \frac{\alpha }{n} \right)^k 
\, \right\} \; , 
\nonumber 
\ena 
where we have used $\bar{P}(0)= \int dx \, P(x) = 1$ and the definition 
of the cumulants of $P$ in (\ref{eq:cl15b}). Using $c_1[P] = p_1 = \bar{x} 
= \bar{y}$, we find: 
\be 
Q_n (y) \; = \;  \int \frac{d\alpha }{2\pi } \; \exp [i \, \alpha \, 
(y - \bar{y} ) ] \; 
\exp \left\{ - \; \frac{1}{2} \; c_2[P] \; 
\left( \frac{\alpha^2 }{n} \right) \; + \; {\cal O}(\alpha^3/n^2)  
\, \right\} 
\label{eq:cl18} 
\en
Note that the dominant contributions from the $\alpha $ integration 
arises from the regime where $\alpha < \sqrt{n}$. In this regime, 
the correction term is of order 
$$ 
{\cal O}(\alpha^3/n^2) \; \approx \; {\cal O}(1/\sqrt{n} )  
$$
and will be neglected for sufficiently large $n$. 
The remaining integral can be easily performed: 
\be 
Q_n (y) \; \approx \; \frac{1}{\sqrt{ 2 \pi } \, \sigma } \; 
\exp \left\{ - \; \frac{ (y - \bar{y} )^2 }{2 \sigma ^2 } 
\right\} \; , \hbo \sigma ^2 \; = \; \frac{c_2[P]}{n} \; . 
\label{eq:cl19} 
\en
which is the celebrated Gaussian distribution. 
This finding is called the {\it central limit theorem:} if the moments 
of probability distribution $P(x)$ exist, the probability distribution 
for the average $y$ can be approximated by a Gaussian for sufficiently 
large $n$ given that the standard deviation $\sigma $ is properly scaled 
with $n$.

\vskip 0.3cm 
\begin{figure}[t]
\centerline{ 
\epsfxsize=10cm
\epsffile{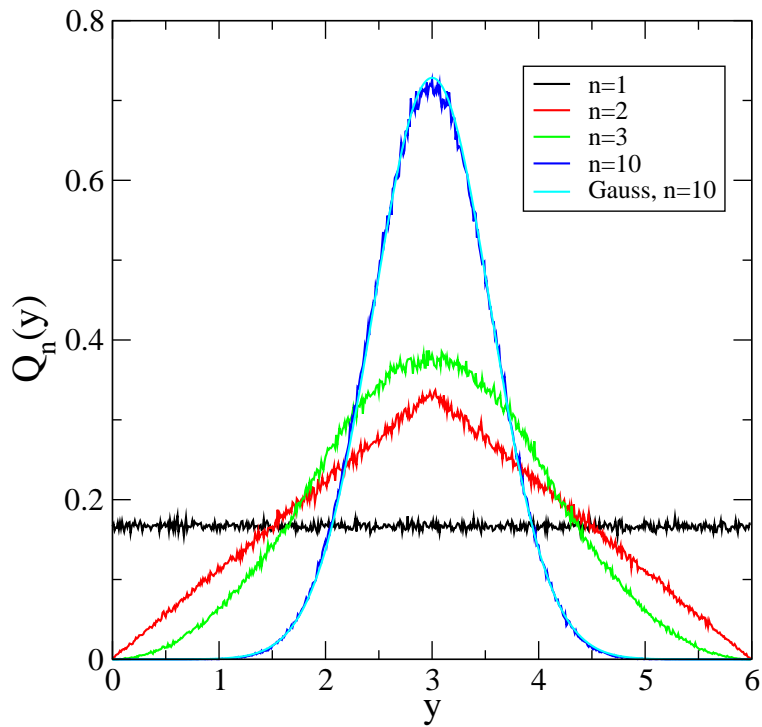}
}
\caption{ Illustration of the central limit theorem: 
probability distributions of the average $y$ after $n$ measurements. 
}
\label{fig:1} 
\end{figure}
Let us discuss this result. Figure~\ref{fig:1} shows the distributions 
$Q_n(y)$ for $n=1,2,3$ and $n=10$. Already for 
$n=10$ the distribution is well approximated by the Gaussian. 

\vskip 0.3cm 
The existence of at least the moment $c_2[P]$ of the single event 
distribution is crucial for the error reduction by repeated 
measurements. Let us consider a Lorentz distribution for the moment:
\be 
P_L(x) \; = \; \frac{1}{\pi b} \; \frac{1}{1 \, + \, (x/b)^2 } \; . 
\label{eq:cl20} 
\en
With the naked eye this distribution resembles the 
Gaussian. The crucial difference is, however, that the second moment 
does not exist: 
$$ 
\int dx \; x^2 \; P_L(x) \; \longrightarrow \; \infty \; . 
$$
The Fourier transform of $P_L(x)$ does, however, exist
\be 
\bar{P}_L(\beta ) \; = \; \exp \Bigl\{ - \, b \, \vert \beta \vert \, 
\Bigr\} \; . 
\label{eq:cl21} 
\en
If we now repeat the measurements $n$ times, the distribution 
of the average $y$ is, according to (\ref{eq:cl8}), by 
$$
Q_n (y) \; = \; \int \frac{d\alpha }{2\pi } \; \exp (i \, \alpha \, y ) \; 
\left[  \exp \left( - b \; \frac{ \alpha }{n} \right) \, 
\right]^n \; = \; P_L(y) \; . 
$$
Obviously, the probability distribution does not change at all even 
if we repeat the measurements many times. This actually implies 
that it impossible to experimentally gain a reliable value for 
the observable $\bar{x}$.

\subsection{Error analysis }

Let us return to the example in (\ref{eq:cl1b}), and let us assume 
a group of experimentalists has performed $n=12$ measurements 
with the result: 
\begin{verbatim} 

            2.813   2.021   0.331   0.865
            5.394   5.937   5.027   1.556
            0.325   2.750   1.283   3.890
\end{verbatim} 
The average over these values and an estimate $\langle x^2 \rangle $ 
of $c_2[P]$ are given by 
\be 
y \; = \; \frac{1}{n} \sum _{k=1}^n x_k \; \approx \; 
2.683 \; , \hbo 
\langle x^2 \rangle \; = \; \frac{1}{n} \sum _{k=1}^n x^2_k \; \approx \; 
3.581 \; . 
\label{eq:cl30}
\en 
We point out 
that $3.581$ is a poor estimate of the true value (\ref{eq:cl17b}) 
of $c_2[P]=3$, but it gives the order of magnitude. 
With this estimate for $c_2[P]$, we find for the error (\ref{eq:cl17}) 
$$ 
\sigma (n=12) \; \approx \; \sqrt{ \frac{3.581}{12} }  \; \approx \; 
0.546 \; . 
$$
Hence, the final 'experimental' result for the observable would be 
\be 
\bar{x} \; \approx \; 2.683  \pm 0.546 \; = \; 2.7(5) \; . 
\label{eq:cl31}
\en 
Note that the true result $\bar{x} = 3$ lies well with the reach of the 
error bars. 

\vskip 0.3cm 
The above experiment was repeated by several research labs. 
Depending on the budget and the focus of research, different 
labs produce different numbers $n$ of measurements: 

\bigskip
\begin{center}
\begin{tabular}{lcccc} 
  & CERN & GSI & DESY & BNL \\ 
$n$ & $120$ & $50$ & $78$ & $150$ \\ 
$y$ & $3.112\pm 0.163 $ & $2.764 \pm  0.255$ & $3.110 \pm 0.207 $& 
$3.083 \pm  0.143$ \\ 
\end{tabular} 
\end{center} 

\bigskip
The smallest error was produced by the largest experiment (BNL). 
We could just quote their result, but it would be a pity to 
disregard a total of $248$ measurements which were carried out 
by the other groups. How can we obtain a \lq world average\rq for the 
observable $\bar{x}$ and how can we quantify its (statistical) error? 

\vskip 0.3cm 
To answer these questions, we assume that the number $n$ of 
each measurement was large enough to approximate the distribution 
of an individual result $y_k$, $k=1 \ldots N$ (where $N=4$ for the 
above example) by a Gaussian (\ref{eq:cl19}): 
\be 
Q (y_l) \; \approx \; \frac{1}{\sqrt{ 2 \pi } \, \sigma _l } \; 
\exp \left\{ - \; \frac{ (y_l - \bar{x} )^2 }{2 \sigma _l ^2 } 
\right\} \; . 
\label{eq:cl32} 
\en
For the world average $y$ we make the ansatz 
\be 
y \; = \; \sum _{l=1}^N a_l \, y_l \; , \hbo 
\sum _{l=1}^N a_l \; = \; 1 \; . 
\label{eq:cl33}
\en 
The weights $a_l$ must be chosen in an optimal way. This choice 
will depend on the errors $\sigma _l$ of the individual experiments. 
In particular, the experiment with the smallest error should contribute 
to the world average with the largest weight. 
Assuming that the experiments at the different labs were 
carried out independently, the probability distribution of the world 
average is now given by 
\be 
W(y) \; = \; \int \prod _{k=1}^N dy_k \; \delta \Bigl( 
y \; - \; \sum _{l=1}^N a_l \, y_l \Bigr) \; 
Q(y_1) \ldots Q(y_N) \; . 
\label{eq:cl34}
\en 
Representing the $\delta $-function in terms of a Fourier integral 
over $\alpha $ (see (\ref{eq:cl6b})), the integrations over 
$y_1 \ldots y_N$ can be easily performed: 
$$ 
W(y) \; = \; \int \frac{d \alpha }{2\pi } \; 
\exp \{ i ( y - \bar{x}) \} \; 
\exp \left\{ - \; \frac{\alpha ^2 }{2} \;  \sum _l a^2_l \sigma _l^2 
\right\} \; . 
$$
Performing the $\alpha $ integration finally fields: 
\be 
Q (y) \; \approx \; \frac{1}{\sqrt{ 2 \pi } \, \sigma  } \; 
\exp \left\{ - \; \frac{ (y - \bar{x} )^2 }{2 \sigma ^2 } 
\right\} \; , \hbo \sigma ^2 \; = \;  \sum _l a^2_l \sigma _l^2 \; . 
\label{eq:cl35} 
\en
The optimal result is achieved if the error squared, i.e., $\sigma ^2$, 
is as small as possible. Here, we must take into account the 
normalisation condition in (\ref{eq:cl33}). We therefore minimise 
\be 
 \sum _l \Bigl[a^2_l \sigma _l^2 \; - \; \lambda \, a_l \Bigr] 
\; \longrightarrow \; \hbox{min} \; , 
\label{eq:cl36} 
\en
where $\lambda $ is a Lagrange multiplier. The global minimum is 
easily obtained: 
\be 
a_l \; = \; \frac{\lambda }{2 \, \sigma _l^2 } \; , \hbo 
\frac{2}{\lambda } \; = \; \sum _l \frac{1}{\sigma _l^2} 
\; . 
\label{eq:cl37} 
\en
The minimal value for $\sigma ^2$ then satisfies
\be 
\sigma ^2 \; = \; \frac{\lambda}{2} \; \hbo \Rightarrow \hbo 
\frac{1}{\sigma ^2 } \; = \; \sum _l \frac{1}{\sigma _l^2} \; . 
\label{eq:cl38} 
\en 
The optimal choice for the weights can therefore also be written as 
\be 
a_l \; = \; \frac{\sigma ^2 }{\sigma _l^2 } \; . 
\label{eq:cl38b} 
\en 
Let us return to the above example. We find 
\bea 
\sigma &\approx & 0.089 \; , 
\label{eq:cl39} \\
a_1  &\approx & 0.30 \; , \hbo a_2 \, \approx \, 0.12 \; , \hbo 
a_3 \, \approx \, 0.19 \; , \hbo a_3 \, \approx \, 0.39 \; . 
\nonumber 
\ena 
With the weights at our disposal, we easily find the optimal value 
for the world average $ y \; \approx \; 3.059 $. 
Together with the error in (\ref{eq:cl39}), the final result is 
\be 
\bar{x} \; = \;  3.059 \, \pm \, 0.089 \; = \; 3.06(9) \; . 
\label{eq:cl40} 
\en
Note that the true result $\bar{x}=3$ is again covered within error bars 
and that the error became significantly smaller than that of 
the best result provided by the BNL group. 

\subsection{Autocorrelations \label{sec:auto} }

In the previous subsections, we have repeatedly assumed that the 
measurements $x_i$ are independent. We will see below that 
a vital tool of computational quantum field theory is to use information 
from the measurement $x_i$ to obtain the value for $x_{i+1}$. 
In this case, the data set $x_i$, $i=1 \ldots n$ is generated by the 
chain 
$$ 
x_1 \, \rightarrow \, x_2  \, \rightarrow \, \ldots \, \rightarrow 
\, x_{n-1} \, \rightarrow \, x_n , 
$$
and the probability of finding a particular set does not factorise 
anymore: 
$$
P(x_1, \ldots , x_n) \; \not= \;  P(x_1) \; \ldots \; P(x_n) \; . 
$$
In the context of QFT simulations we will, however, make an effort 
to render the values $x_i$ as independent as possible. 
This generically implies that events which are separated by 
some \lq time\rq $\tau $, i.e., the events $x_i$ and $x_{i+\tau }$, 
can be considered as statistically independent. 
The trick for obtaining an idea of the error of the estimator is 
to group $b$ measurements together: 
\be 
y_\nu \; = \; \frac{1}{b} \; \sum _{i=1}^b x_{\nu + i} \; , \hbo 
\nu = 1 \,  \ldots \, M \; , \hbo M \, = \, \frac{n}{b} \; , 
\label{eq:cl50}
\en 
where we choose 
\be 
1 \; \le \; \tau \; \ll \; b \; .  
\label{eq:cl51}
\en 
This latter constraint implies that the values $y_\nu$ are statistically 
independent and that they have a Gaussian distribution because 
of the central limit theorem. The quantities of interest are 
the average 
\be 
\bar{y} \; = \; \frac{1}{M} \sum _{\nu =1}^M y_\nu \; , 
\label{eq:cl52}
\en 
which converges to the observable $\bar{x}$ in the limit $M \to \infty$, 
and the corresponding error 
\be 
\sigma ^2 \; = \; \frac{1}{M} \; c_2[P_y] \; = \; \frac{b}{n} \; 
c_2[P_y] \; . 
\label{eq:cl53}
\en
where the cumulant $c_2[P_y]$ is given by 
\be 
c_2[P_y] \; = \; \left\langle 
\frac{1}{M} \sum _{\nu =1 }^M y_\nu ^2 \; \right\rangle  \; 
\; - \; 
\left[ \left\langle  
\frac{1}{M} \sum _{\nu =1 }^M y_\nu \; \right\rangle  \; \right]^2 \, , 
\label{eq:cl54}
\en
with 
$$ 
\langle f \rangle \; = \; \int \prod _{l=1}^n dx_l \; f(x_1 \ldots x_n) \; 
P(x_1, \ldots , x_n) \; . 
$$
Assuming translational invariance, i.e., 
$$ 
\langle x_{\nu + i} x_{\nu + k} \rangle \; = \; 
\langle x_{i} x_{ k} \rangle \; , 
$$
we find 
\be 
c_2[P_y] \; = \; \frac{1}{b^2} \sum _{i=1}^b \sum _{k=1}^b c(k-i) \; , \hbo 
c(k-i) \; = \; \langle x_{i} x_{ k} \rangle \; - \; 
 \langle x_{i}  \rangle  \langle x_{ k} \rangle \; ,  
\label{eq:cl55}
\en
where $c(k-i)$ is called the {\it autocorrelation function}. 
Introducing the relative distance $t = k-i$, and trading the summation 
over $k$ in (\ref{eq:cl55}) for a summation over $t$, we find
\be 
c_2[P_y] \; = \; \frac{1}{b^2} \sum _{i=1}^b \sum _{t=1-i}^{b-i} c(t) \; . 
\label{eq:cl56}
\en
Interchanging the summation indices $t$ and $i$ and after summing over $i$, 
this last equation becomes:
$$
c_2[P_y] \; = \; \frac{1}{b} c(0) \; + \; \frac{2}{b^2} \, 
\sum _{t=1}^{b-1} (b-t) \; c(t) \; . 
$$
We have already mentioned that the measurements $x_i$ and $x_{i+\tau}$ 
are (almost) uncorrelated. The equivalent statement is that the 
correlation function vanishes for sufficiently large arguments: 
$$
c(t) \; \approx \; 0 \; , \hbo \hbox{for} \; \; \, t > \tau \; . 
$$
For $b \gg \tau $, we approximately find: 
\be 
c_2[P_y] \; \approx \;  \frac{1}{b} c(0) \; + \; \frac{2}{b^2} \, 
\sum _{t=1}^{b-1} b \; c(t) \; = \; \frac{1}{b} 
\sum _{t=1-b}^{b-1} c(t) \; . 
\label{eq:cl57}
\en
It is convenient to introduce the {\it normalised autocorrelation 
function } by 
\be 
\rho (t) \; = \; \frac{c(t) }{ c(0) } \; , \hbo 
c(0) \; = \; \langle x^2 \rangle - \langle x \rangle ^2 \; = \; 
c_2[P_x] \; . 
\label{eq:cl58}
\en
The {\it integrated autocorrelation time } is then defined by 
\be 
\tau \; = \; \frac{1}{2} \sum _{t=1-b}^{b-1} \rho (t) 
\; = \; \frac{1}{2} \; + \; \sum _{t=1}^{b-1} \rho (t) \; . 
\label{eq:cl59}
\en
Inserting (\ref{eq:cl59},\ref{eq:cl58},\ref{eq:cl57}) into 
(\ref{eq:cl53}), we finally obtain for the error which should be attributed 
to our estimate $\bar{y}$ in (\ref{eq:cl52}): 
\be 
\sigma ^2 \; = \, \frac{2 \, \tau }{n} \; c_2 [P_x] \; . 
\label{eq:c60}
\en
Let us perform a consistency check by considering the special case that 
the measurements are uncorrelated. In this case, the autocorrelation 
function $\rho(t)$ vanishes for $t\ge 1$, and the autocorrelation time 
is given by $\tau = 1/2$. We indeed recover the familiar result 
\be 
\sigma ^2 \; = \, \frac{1}{n} \; c_2 [P_x] \; , \hbo 
\hbox{for independent measurements.} 
\label{eq:c61}
\en
Note that autocorrelations increase the error bars. Not knowing 
the autocorrelations in a numerical simulations leads us to the 
erroneous assumption that the error of the estimator is given by 
(\ref{eq:c61}), while the true result is by a factor $\tau $ larger. 
Not knowing the autocorrelations always leads to an underestimation 
of the statistical error.

\section{Lessons from the Ising model }
\subsection{Phase transitions }

A phase transition occurs if the properties of matter change 
qualitatively when an external parameter, such as temperature, is 
altered. The phase transition of water from a liquid to a gas 
phase when the temperature exceeds roughly  $\approx 100^0\,$ Celsius 
(under normal conditions), is well known from everyday life. 
A second example is the ferromagnet: the interaction between 
microscopic spins favour a unique orientation of the spins. 
This yields an {\it ordered phase} at low temperatures. Above 
the critical temperature, called the {\it Curie temperature } in the 
present context, the spins are organised in a {\it disordered phase}. 

\vskip 0.3cm 
Let us assume that a ferromagnet is in the disordered phase 
at a temperature slightly bigger than the critical temperature 
$T_c$. If we decrease the temperature, the information of the 
unique orientation spreads over the spin lattice. 
This \lq Gedankenexperiment \rq shows that the spatial correlation 
of the spins becomes large at the critical temperature. 
This phenomenon, is quantified with the help of the spin-spin 
{\it correlation function}: 
\bea 
\biggl\langle \sigma (x) \, \sigma (y) \biggr\rangle \; \propto \; 
\exp \left\{ - \frac{ \vert x-y \vert }{ \xi } \right\} \; . 
\label{kl:1} 
\ena 
The {\it correlation length } $\xi $ obviously measures the 
spatial distance over which the spins show roughly the same 
orientation. 
Close to the phase transition, i.e.~for $T \stackrel{>}
{_\sim}T_c$, $\xi $ becomes large anticipating the ordered phase: 
\be 
\xi \; \approx \; \xi_+ \, \left\vert 1 - \frac{T}{T_c} \right\vert ^{- \nu } 
\; , \hbox to 1cm {\hfill } 
(T \stackrel{>}{_\sim}T_c) \; , 
\label{kl:2} 
\en
where $\nu $ is a positive number. The divergence of the correlation 
length at the phase transition is characteristic for 
a transition of {\it 2nd } (or higher) order. 
In the case of a {\it 1st order} transition, the increase of $\xi $ 
is hindered by the nucleation of bubbles which contain 
chunks of the new state of matter. These bubbles provide additional 
disorder and the correlation length stays finite. 

\vskip 0.3cm 
For phase transitions above first order, the 
singularity of the correlation length has its fingerprint 
in many other thermodynamical quantities such as the {\it specific heat} 
$C$ or the {\it magnetic susceptibility} $\chi $:
\bea 
C \; \approx \;  C_0 \, \left\vert 1 - \frac{T}{T_c} \right\vert ^{- \alpha} 
\; , \hbox to 1cm {\hfill } 
\chi & \approx & \chi _0 \, \left\vert 1 - \frac{T}{T_c} \right\vert 
^{- \gamma } \; , \hbox to 1cm {\hfill } 
(T \stackrel{>}{_\sim}T_c) \; . 
\nonumber 
\ena
The {\it critical exponents } $\nu $, $\alpha $, $\gamma $ 
are independent of the microscopic properties of the spin model 
(such as the lattice geometry), and only depend on the symmetries 
(present at the transition) and the number of dimensions. 
They are often used to sort solid state physics models into the so-called 
{\it universality classes}.

\subsection{Quantum field theory rising }

Let the {\it lattice spacing} $a$ denote the distance between two 
neighbouring lattice sites . In the previous subsection, we found that 
the correlation length diverges if the coupling constant $\beta $ 
(or inverse temperature in the present context, $\beta = 1/T$) approaches 
its critical value (see  (\ref{kl:2})). This statement can be 
phrased in units of the lattice spacing as 
\be 
\frac{\xi }{a} \; = \; \kappa \, \biggl( \beta _c \, - \, \beta 
\biggr) ^{-\nu } \; , \hbo \beta \stackrel{<}{_\sim } \beta _c \; , 
\label{eq:41} 
\en 
where $\kappa $ is a dimensionless constant which can be obtained 
by numerical means. 

\vskip 0.3cm 
Let us now reinterpret these findings. 
Rather than saying that $\xi $ diverges and $a$ is fixed, we say that 
the correlation length $\xi $ is fixed and is given by an observable 
in physical units. 
We will see that this interpretation of the same data defines 
a {\tt quantum field theory}. For fixed correlation length $\xi $, 
(\ref{eq:41}) defines the lattice spacing as a function of $\beta $, 
i.e., $a \to a(\beta )$, 
\be 
a(\beta ) \; = \; \frac{1}{\kappa } \; ( \beta - \beta _c)^\nu \; \xi \; , 
\hbo \beta \rightarrow \beta _c \; .  
\label{eq:42} 
\en 
The key point is if we make the number of spins bigger and bigger and, 
at the same time, the distance $a$ between the spins smaller and smaller, 
we will obtain a field theory in the limit $a \to 0$. 
For the 2d Ising model on a cubic lattice, we have $\beta _c \approx 0.44$ 
and $\nu =1$. The field theory limit is then approached 
when $a$ vanishes linearly with $\beta - \beta _c$: 
$$ 
a(\beta ) \; = \; \frac{\xi }{\kappa } \; ( \beta - \beta _c) \; , 
\hbo \hbox{(2d Ising model).}
$$

\begin{figure}[t]
\centerline{ 
\epsfxsize=10cm
\epsffile{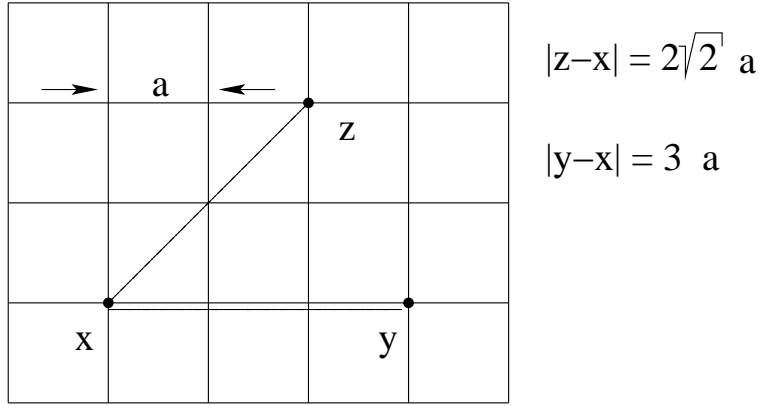}
}
\caption{ Spin correlation along the diagonal and the symmetry axis, 
   respectively. } 
\label{fig:4} 
\end{figure}
\vskip 0.3cm 
Note that the dimensionless parameter $\beta $ is not at our disposal 
anymore, since it specifies the magnitude of the lattice spacing. 
Instead, the value of $\xi $ parameterises the
emerging quantum field theory. The exchange of a dimensionless parameter 
for a scale dependent one in a quantum field theory 
is known as {\it dimensional transmutation.} It is a generic 
feature of quantum field theories. For instance in the case of 
perturbative QCD, the dimensionless gauge coupling $g$ is eliminated 
in favour of the scale dependent parameter $\Lambda _{QCD}$. 

\vskip 0.3cm 
Let us assume that a certain correlation function was obtained 
by a numerical simulation of a classical lattice model for large 
values $\vert  x -y \vert $, 
\be 
D\biggl(\vert x -y \vert \biggr) \; = \; 
\biggl\langle F(\phi(x)) \; F(\phi(y)) \biggr\rangle \; \propto \; 
\exp \biggl\{ - m  \, \vert x -y \vert \biggr\} \; , 
\label{eq:42a} 
\en 
where $m$ is called the {\tt screening mass}. Since the distance 
$ \vert x -y \vert $ is only known in units of the lattice 
spacing by construction, the simulation will provide the 
mass in units of the lattice spacing as a function of $\beta $, i.e. 
$m a \, (\beta )$. If universality holds, 
one finds the characteristic scaling of the lattice model, i.e., 
\be 
m \, a(\beta ) \; = \; \kappa _m \; \, \biggl( \beta _c \, - \, \beta 
\biggr) ^{\nu } \; , \hbo \beta \stackrel{<}{_\sim } \beta _c \; . 
\label{eq:43} 
\en 
Using (\ref{eq:41}), we see that the product  $m \, \xi $ approaches a 
constant in the vicinity of the critical limit: 
\be 
 m \, \xi  \; = \;  m \, a \, \frac{ \xi }{ a } \; 
= \;  \kappa _m \,  \kappa \; . 
\label{eq:44} 
\en 
Note that $\kappa $ and $\kappa _m$ are two c-numbers which we obtain
from the numerical simulations. With the help of these two 
numbers we can ``measure'' the desired mass $m$ in units of 
$1/\xi $, where $\xi $ is the only free parameter of our theory. 

\vskip 0.3cm 
In the case of a quantum field theory, we expect that due to the 
isotropy of the vacuum the correlation function (\ref{eq:42a}) 
only depends on the distance between $x$ and $y$. In the classical 
lattice model, continuous rotational symmetry is violated due to the 
presence of the cubic lattice, and it might happen that the 
quantum field theory which emerges from the lattice model inherits the 
anisotropy. This anisotropy can be measured by comparing the 
correlation length in lattice units along a lattice symmetry axis, 
$\xi $, and a long the diagonal direction, $\xi _d$ (see figure 
\ref{fig:4}). As far as global symmetries are concerned, the symmetry 
under consideration is restored in the critical limit (\ref{eq:42}): 
$$
\xi _d \; = \; \xi \; , \hbo \hbox{for} \; \; a \to 0 \; . 
$$
Further details on the restoration of rotational symmetry in the 
context of the 2-dimensional Ising model can be found in~\cite{bell}.  

\subsection{Mean-field approximation} 

The starting point for a thermodynamical description of the Ising model 
is the partition function: 
\bea 
{\cal Z} = \sum_{\{ \sigma _x \}} {\exp\left(- \beta \, H(\sigma ) \right)} \, 
\label{kl:3} 
\ena 
where $\beta = 1/T$ and where a {\it spin } $\sigma _x = \pm 1$ 
is associated which each site $x$ of the square lattice. The sum in 
(\ref{kl:3}) extends over all possible spin configurations. 
The {\it ferromagnetic } interaction favours a unique orientation of the 
spins, and is described by 
\bea 
H(\sigma ) =  - \sum_{<xy>} \sigma _x \sigma _y \; , 
\label{kl:4} 
\ena
where the sum extends over all pairs $<xy>$ of nearest neighbours. 
In order to preserve translational invariance, periodic 
boundary conditions are often used in particle physics applications, 
although these conditions are difficult to interpret in the solid state 
physics context. 

\begin{figure}[t]
\centerline{ 
\epsfxsize=12cm
\epsffile{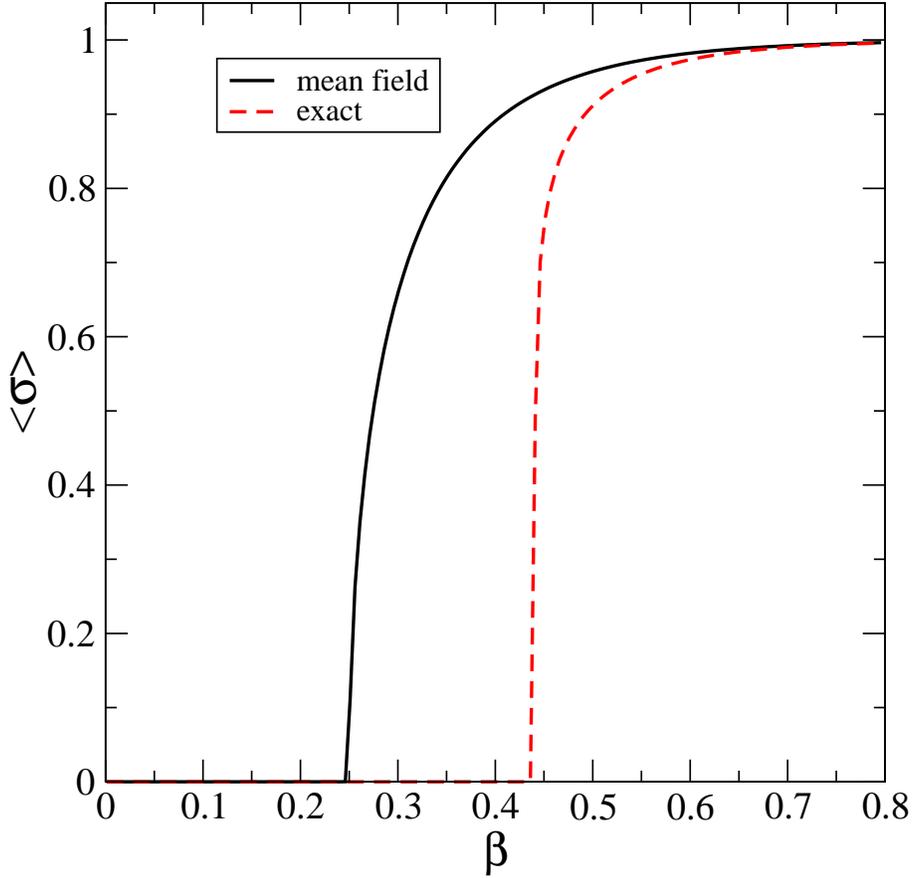}
}
\caption{ Magnetisation per site as function of the inverse 
temperature $\beta $; solid line: mean field approximation; 
dashed line: exact } 
\label{fig:mf1} 
\end{figure}
In order to gain an initial insight into the phase structure of the 
Ising model, we choose a particular spin $\sigma _{x_0}$ of the lattice, and 
assume heuristically that that we might replace the neighbouring spins 
by the mean 
\be 
\langle \sigma \rangle \; = \; \frac{1}{\cal Z} \; 
\sum_{\{ \sigma _x \}} \; \sigma _{x_0} \; 
\exp\left(- \beta \, H(\sigma ) \right) \, . 
\label{eq:mf1} 
\en 
The Hamiltonian is then approximately given by 
\be 
H(\sigma _{x_0} ) \; \approx \; \hbox{const.} \; - \; 
4 \langle \sigma \rangle \;  \sigma _{x_0} \; . 
\label{eq:mf2} 
\en 
Note that each spin possesses $4$ neighbours on a cubic 2d square lattice. 
Equation (\ref{eq:mf1}) turns into a self-consistency equation 
to determine the $\langle \sigma \rangle $, which can be 
interpreted as the magnetisation per site: 
\bea 
\langle \sigma \rangle &=& \frac{1}{\cal N } \; 
\sum_{\sigma  _{x_0} = \pm 1}  \; \sigma  _{x_0} \; 
\exp\left(- \beta \, H(\sigma ) \right) \, , 
\label{kl:mf3} \\
{\cal N } &=& \sum_{\sigma  _{x_0} = \pm 1}  \; 
\exp\left(- \beta \, H(\sigma ) \right) \, . 
\label{kl:mf4} 
\ena
Performing the sum over $\sigma  _{x_0}$ leaves us with a non-linear 
equation: 
\be 
\langle \sigma \rangle \; = \; \hbox{tanh} \, \Bigl( 
4 \, \beta \, \langle \sigma \rangle \Bigr) \; . 
\label{kl:mf5} 
\en
Before we proceed with a numerical solution of this  
equation, we point out that (\ref{kl:mf5}) always possesses the 
trivial solution 
$$ 
\langle \sigma \rangle \; = \;0 \; . 
$$
A graphical inspection of (\ref{kl:mf5}) easily shows that for 
\be 
\beta \; > \; \frac{1}{4} \; , 
\label{kl:mf6} 
\en
two non-trivial solutions $\pm c$, for $c>0$ exist. 
The physical interpretation of the solution is clear: for 
sufficiently small temperature (high $\beta $), an ordered phase 
exists. The critical value is, in mean-field approximation, given by 
\be 
\beta _c^\mathrm{MF} \; = \; 1/4 \; . 
\label{kl:mf7} 
\en
Equation (\ref{kl:mf5}) can be easily solved numerically 
with the Newton method or by fixed point iteration. 
The result for the magnetisation as a function of the 
inverse temperature is shown in figure~\ref{fig:mf1}. 
Also shown is the exact result~\cite{Kramers:1941kn,Yang:1952xj}: 
\be 
\langle \sigma \rangle \; = \; \left[ 1 \; - \; 
\frac{1}{\mathrm{sinh} ^4 (2\beta ) } \right] ^{1/8} 
\label{kl:mf7b} 
\en
The mean-field result qualitatively reproduces the correct 
phase structure. The mean field approximation is able to 
describe the transition from the disordered to the ordered phase. 
However, the mean field approximation fails at a quantitative 
level. The correct value for the critical value, which was already 
obtained by Kramers and Wannier in 1941~\cite{Kramers:1941kn}, is given 
by 
\be 
\beta _c \; = \; 0.44068679 \ldots \hbo 
\label{kl:mf8} 
\en
is significantly underestimated. Also the rise of the magnetisation 
close to $\beta _c$ is not correctly reproduced. 
A Taylor expansion of (\ref{kl:mf5}) with respect to $\beta $ around 
$\beta _c^\mathrm{MF} =1/4$ (and therefore also with respect to $\sigma $), 
yields: 
\be 
\langle \sigma \rangle \; \approx \; \sqrt{12} \; \Bigl( \beta - \beta_c 
\Bigr)^b \; , \hbo b = \frac{1}{2} \; . 
\label{kl:mf9} 
\en
The mean field critical exponent of $1/2$ is much too large compared with 
the exact exponent of $b_{exact} = 1/8$. 

\vskip 0.3cm 
The advantage of the mean-field approximation is that it can 
be easily applied to a variety of models (e.g.~the Ising model 
in $d>2$ where no exact results are available). It often 
provides a correct first impression of the phase structure. 
The disadvantage is that it is difficult to improve the 
approximation in a systematic way.

\subsection{Duality transformation \label{sec:dual} }

\begin{figure}[t]
\centerline{ 
\epsfxsize=\textwidth
\epsffile{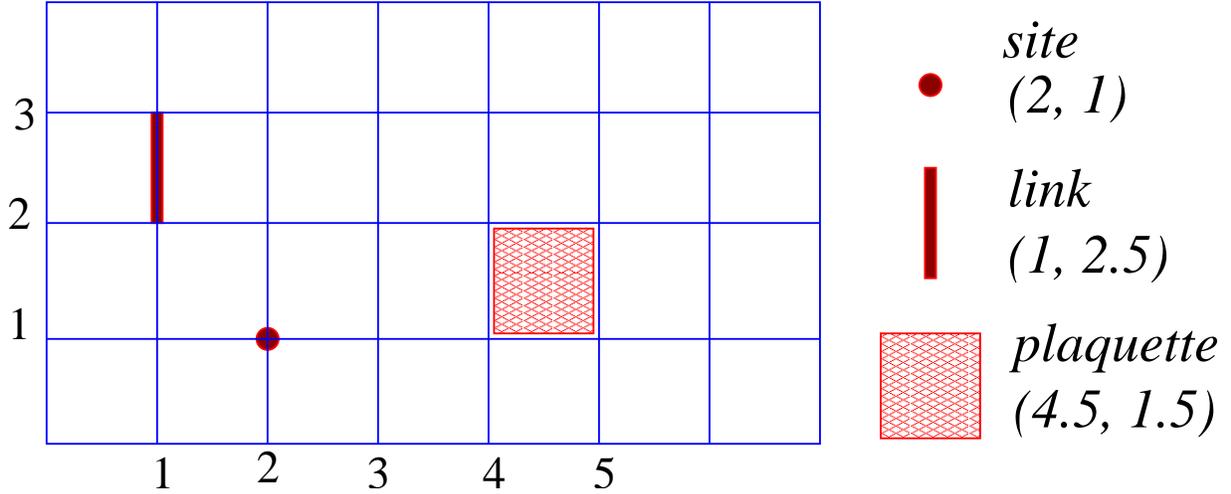}
}
\caption{ Geometrical objects on a lattice. }
\label{fig:geom1} 
\end{figure}
Let us list different geometrical objects on a lattice. 
The {\it sites} on a lattice are labelled by integer coordinates. 
{\it Links} are short line segments which join two neighbouring 
sites on the lattice. In order to unambiguously address a link on the 
lattice, we use coordinates which are integers with the exception 
of one coordinate which is half integer, such as $2.5$ (see figure 
\ref{fig:geom1} for an illustration). 
Another important object is the so-called {\it plaquette }, which 
is an elementary square of the cubic lattice. Two coordinates are 
half integer when a plaquette is addressed. In higher dimensions, there 
are also {\it cubes}, and their coordinates are half integer, 
while the other coordinates are integer. 

\vskip 0.3cm 
The {\it dual lattice} is an important object which helps 
to gain non-perturbative information for certain lattice models. 
The coordinates of the dual lattice are obtained by adding $0.5$ 
to all coordinates of the lattice. If we consider a $d$ dimensional 
lattice model, the duality transformation maps an $x$-dimensional 
geometrical object into a $d-x$ dimensional object on the dual lattice. 
Let us consider $2$ dimensions. A site, such as $(2,4)$ is mapped 
into $(2.5, \, 4.5)$, which are the coordinates of a plaquette, while 
a link, e.g.~$(1.5, \, 5)$, maps into another link namely 
$(2,\, 5.5)$. 

\vskip 0.3cm 
With these prerequisites, let us consider the probabilistic measure 
of the 2d Ising model. Since the product $\sigma _x \sigma _y$ can 
only be $\pm 1$, we expand: 
$$ 
\exp \Bigl\{ \beta \, \sigma _x \sigma _y \Bigr\} \; = \; 
a \; + \; b \;  \sigma _x \sigma _y \; . 
$$
Inserting both possible values for the product $\sigma _x \sigma _y$, we find: 
$$ 
a \; + \; b \; = \; \mathrm{e}^\beta \; , \hbo 
a \; - \; b \; = \; \mathrm{e}^{ - \beta } \; , 
$$
and finally: 
\be 
\exp \Bigl\{ \beta \, \sigma _x \sigma _y \Bigr\} \; = \; 
\cosh \beta \; + \; \sinh \beta \; \sigma _x \sigma _y \; . 
\label{eq:g1}
\en 
Hence, the partition function in (\ref{kl:3}) can be written as 
\be 
{\cal Z} = \sum_{\{ \sigma _x \}} 
\cosh ^{2N} \, \prod _{\langle xy\rangle} \, 
\Bigl[ 1 \; + \; \tanh \beta \;  \sigma _x \sigma _y  \Bigr] \; . 
\label{eq:geom2}
\en 
where $x$ and $y$ are nearest neighbours on the lattice, and the 
corresponding link is denoted by $\langle xy \rangle$. Note also that, 
in $2$ dimensions, there are $2 N$ links for a lattice with $N$ sites. 
In order to perform the sum over all spin configurations in (\ref{eq:geom2}), 
we use the important relations: 
$$ 
\sum _{\sigma = \pm 1} \sigma \; = \; 0 \; , \hbo 
\sum _{\sigma = \pm 1} \sigma ^2 \; = \; 2 \; . 
$$
Hence, if we perform the sum over the spin $\sigma _x$ in  (\ref{eq:geom2}), 
we must make sure that it appears twice (or an even number of times) 
when we expand the products of the square brackets. 
\vskip 0.3cm 
\begin{figure}[t]
\centerline{ 
\epsfxsize=\textwidth
\epsffile{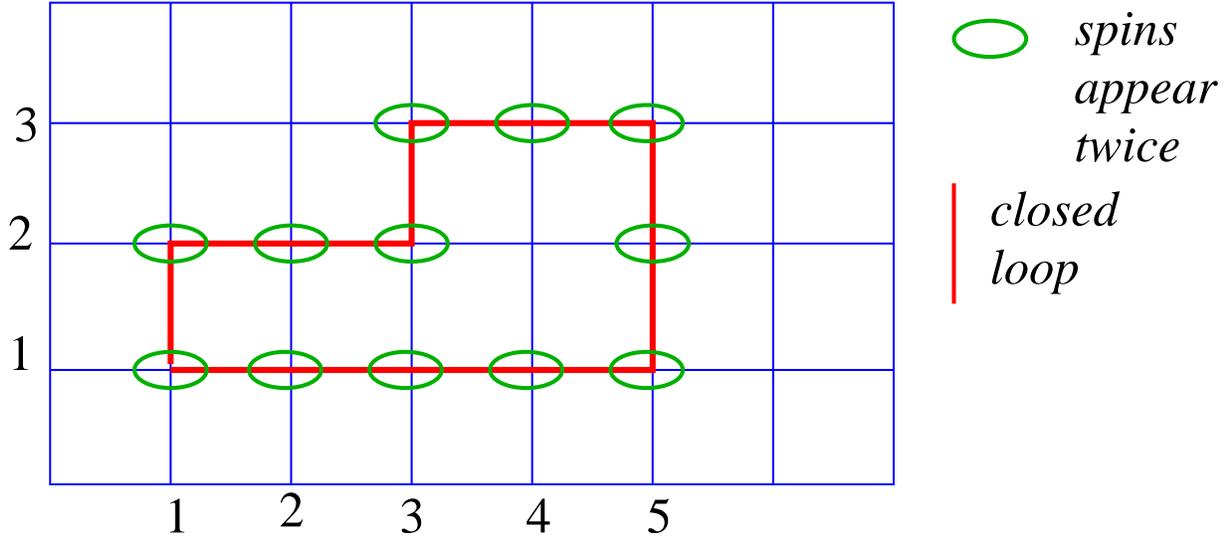}
}
\caption{ Integration over spins generate closed loops.} 
\label{fig:geom2} 
\end{figure}
\vskip 0.3cm 
\begin{figure}[t]
\centerline{ 
\epsfxsize=\textwidth
\epsffile{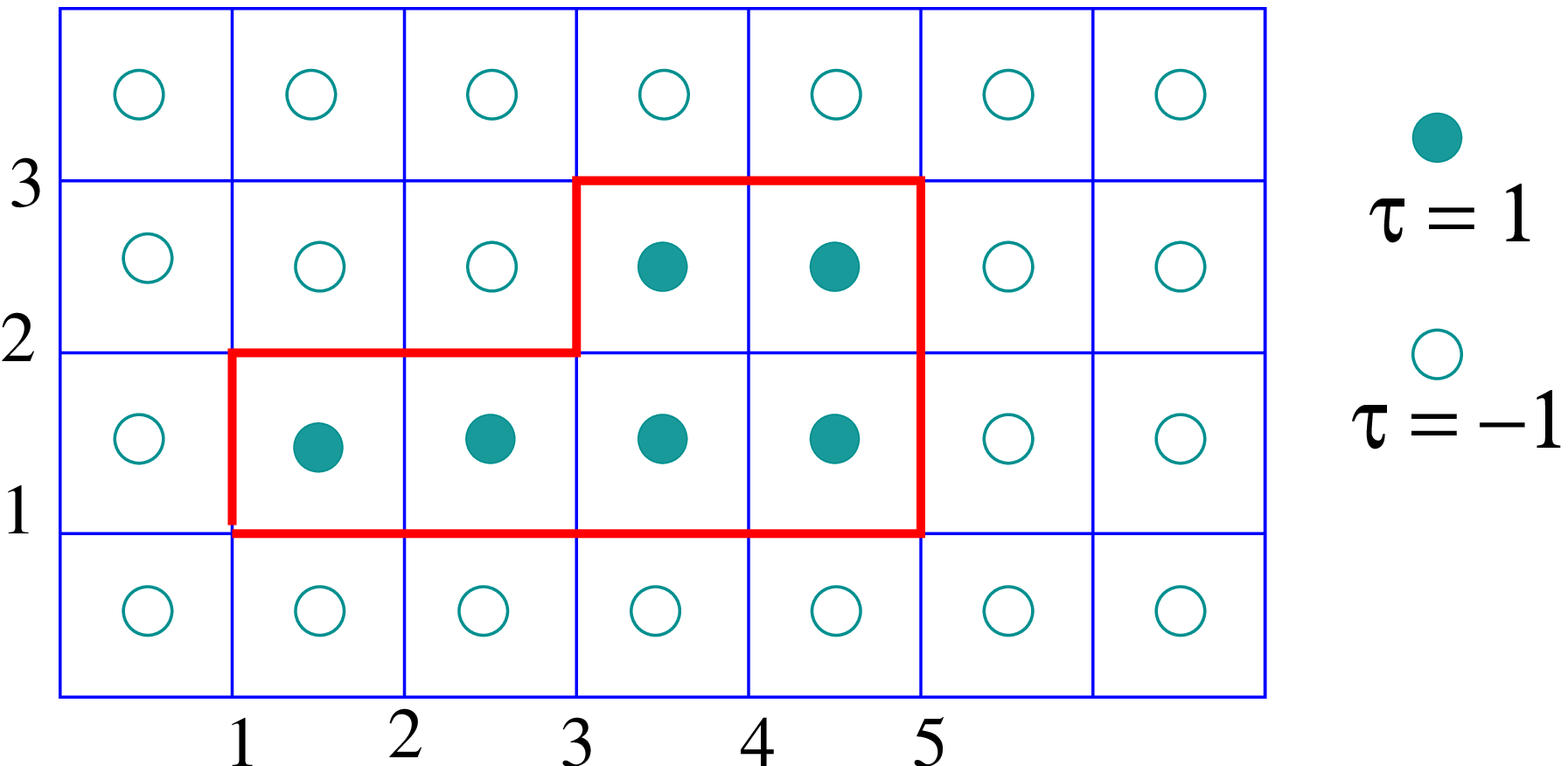}
}
\caption{ Introducing dual variables to represent closed loops. }
\label{fig:geom3} 
\end{figure}
Thus, if we avoid a vanishing contribution to the partition function, 
integration over the spins generates closed loops  the corners 
of which are marked by a pair of spins. For each link of the 
closed loop, we get a factor $\tanh \beta $. Hence, after we have integrated 
out all spins, the partition function can be written as: 
\be 
{\cal Z} = \cosh ^{2N} \beta \; 2^N \; \sum_{\mathrm{loops}} 
\Bigl[ \tanh \beta \;  \Bigr] ^{N(L)} \; , 
\label{eq:geom3}
\en 
where $N(L)$ is the number of links of the closed loop $L$. 
Note that we obtained a factor of $2$ for each sum over a particular spin 
$\sigma $. This gives rise to the prefactor $2^N$ in front of the sum 
in (\ref{eq:geom3}). We now 
have converted the Ising model into a string theory, but we have not 
gained much information on the Ising model so far. 
To proceed further, we must control the sum over closed loops. 
For this purpose, we introduce new variables $\tau = \pm 1 $  which 
are associated with the plaquettes (see figure~\ref{fig:geom3}). 
If we consider two neighbouring plaquettes, there is always  just 
one link between them. Now we say that if the product of the 
two neighbouring plaquettes is $-1$, the corresponding link 
is part of the loop. If the product is $1$, the link is not part of the 
loop. The advantage of the $\tau $ variables is that we can randomly 
assign $\pm 1$ to them and all loops which we produce are closed. 
Hence, summing over all possible $\tau $ configurations will do the 
sum over all possible closed loops for us. 

\vskip 0.3cm 
Note that each plaquette of the lattice is mapped to a site on the 
dual lattice. The link between two neighbouring plaquettes is mapped 
into the link between the adjacent sites of the dual lattice. 
Finally, we must express $N(L)$ in terms of the $\tau $ variables. 
For this purpose, we have to count all {\it activated} links (links which 
are part of a loop) on the lattice. It is easy to check that 
\be 
N(L) \; = \; \sum _{\langle x_d y_d \rangle } 
\frac{1}{2} \; 
\Bigl[ 1 \; - \; \tau _{x_d} \, \tau _{y_d} \, \Bigr] 
\label{eq:geom4}
\en 
counts these links: if two neighbouring $\tau $s are equal, they do not 
contribute to $N(L)$, and is they are different, they contribute $1$ as they 
should. Using the $\tau$-representation of the closed loops, the 
partition function (\ref{eq:geom3}) becomes 
\be 
{\cal Z} = [\cosh  \beta ]^{2N} \, [ 2 \, \tanh \beta ] ^{N} 
\sum_{\{ \tau _{x_d} \}} 
\prod _{\langle  x_d y_d \rangle} \, 
\Bigl[  \tanh \beta \Bigr] ^{- \frac{1}{2}  \tau _{xd} \tau _{yd}  } \; . 
\label{eq:geom5}
\en 
This last equation can be written as 
\bea 
{\cal Z} &=& \sinh ^N (2 \beta )  
\sum_{\{ \tau _{x_d} \}} 
\; \exp \Bigl\{ \sum _{\langle  x_d y_d \rangle} 
\widetilde{\beta } \;  \tau _{xd} \tau _{yd}  \Bigr\} \; . 
\label{eq:geom6} \\ 
\widetilde{\beta } &=& - \, \frac{1}{2} \, \ln \; \tanh \beta \; . 
\label{eq:geom7} 
\ena 
We have obtained again a 2d Ising model which is now formulated on the
dual lattice: the only difference is that the coupling is now 
$\widetilde{\beta }$ rather than $\beta $. 
It is not generally true that the duality transform yields the same 
lattice model just with different couplings. Models which {\it do } have 
this property are called {\it self dual}. 

\vskip 0.3cm 
\begin{figure}[t]
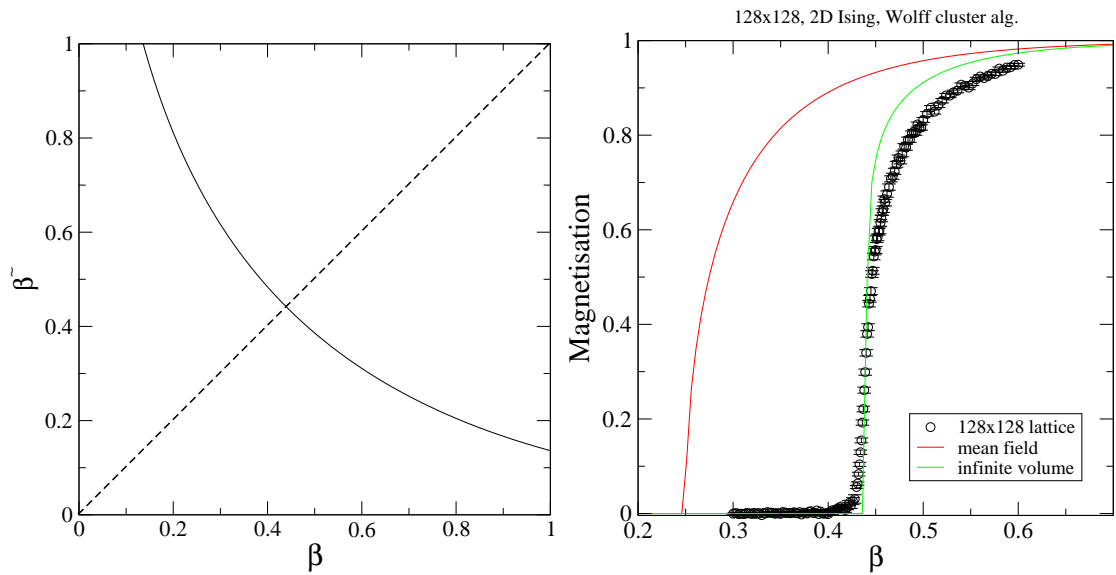

\centerline{ 
\epsfxsize=.45\textwidth
\epsffile{beta.eps}
\epsfxsize=.45\textwidth
\epsffile{mag_cluster2.eps}
}
\caption{ The dual coupling constant $\widetilde{\beta }$ 
as a function of $\beta $ (left). Magnetisation as a function of $\beta $ 
(right). 
}
\label{fig:is1} 
\end{figure}
Now let us assume that $\beta $ is large (small temperature). In this case, 
we find from (\ref{eq:geom7}) that 
$$ 
\widetilde{\beta } \; \approx \; \mathrm{e}^{-2 \beta } \; , \hbo 
\beta \; \hbox{large}. 
$$
By contrast, if $\beta $ is small (the high temperature limit), we find 
$$ 
\widetilde{\beta } \; \approx \; - \frac{1}{2} \, \ln \beta  \; , \hbo 
\beta \; \hbox{small}. 
$$
Hence, large $\beta $ corresponds to small $\widetilde{\beta }$ and vice
versa (see figure~\ref{fig:is1}). 
This is interesting since the so-called {\it strong coupling 
expansion} techniques are available for small $\beta $. 
Performing the expansion with respect to $\widetilde{\beta } $ in the dual 
model, the large $\beta $ regime can also be studied by analytic 
methods. The basis of this expansion is a Taylor expansion of the 
exponential with respect to $\beta $. This expansion naturally 
reaches its radius of convergence when $\beta $ approaches the 
critical coupling $\beta _c$. Performing the expansion using the 
dual model, the Taylor expansion with respect to $\widetilde{\beta }$ 
also breaks down at the critical coupling. There are no other couplings 
for which singularities in thermodynamical quantities occur. 
Hence, the critical point is obtained if 
\be 
\beta \; = \; \widetilde{\beta } \; = \, \beta_c \; . 
\label{eq:geom8} 
\en
Using (\ref{eq:geom7}), we therefore find 
\be 
\beta _c \; = \; - \, \frac{1}{2} \, \ln \; \tanh \beta _c \; , \hbo 
\beta _c \; = \; \frac{1}{2} \, \ln (1 + \sqrt{2}) \; \approx \; 
0.44068679 \ldots \; . 
\label{eq:geom9} 
\en 
Figure~\ref {fig:is1} also shows the magnetisation as a function of 
$\beta $ for a $128 \times 128$ lattice compared with 
the mean field result and the exact result in the infinite volume limit.

\vskip 0.3cm
There are lot of interesting features of field theories 
already present in the Ising model: there is the relation  between 
a lattice model and a theory of strings, and there is the duality transform 
which maps the high temperature theory onto a low temperature theory.

\section{Markov chain Monte-Carlo: the Ising case study }

\subsection{Foundations }
The idea central to all simulations of lattice  models is 
to generate lattice configurations $\{ \sigma _x \}$ 
according to their probabilistic measure 
\bea 
P(\sigma ) \; = \; \exp\left(- \beta \, H(\sigma ) \right) 
/ {\cal Z} 
\label{eq:mc1} 
\ena 
where $ {\cal Z} $ is the partition function (\ref{kl:3}). 
A straightforward idea to accomplish this task would be 
to generate randomly the spins at each site $x$ and to accept or reject 
this configuration according to (\ref{eq:mc1}). 
The problem is that we would hardly find any acceptable configurations. 
Why is this so? 

\vskip 0.3cm 
Let us answer this question in the context of the Ising model of 
the previous section. The two dimensional lattice consists of 
$N \; = \; 125 \times 125 $ sites. Since $\sigma \in \{-1,+1\}$, there 
are $2^N \approx 10^{4704}$ different lattice configurations. 
We further introduce the average action per site, i.e. 
\be 
\bar{s} \; = \;  \frac{1}{N} \biggl\langle \sum _{<xy>} 
\sigma _x \sigma _y \biggr\rangle \; = \; 
\frac{1}{N} \biggl\langle \sum _{x} 
\biggl\langle {\cal A} (x) \biggr\rangle \; = \; 
\biggl\langle {\cal A} (x) \biggr\rangle \; =: \; \bar{\cal A} 
\label{eq:mc2} 
\en 
where 
\be 
{\cal A}(x) := 
\sum _{y>x, \vert x - y \vert =1 } \sigma _x \sigma _y \; , 
\hbo 
\sum _{y>x, \vert x - y \vert =1 } 1 \; = \; 2 \; , 
\label{eq:mc3} 
\en 
and where we have used translational invariance. 
A measure for the strength of the fluctuations of the 
action around its average value $N \, \bar{s}$ is given by 
\bea 
\delta ^2 &=& 
\left\langle \left( \sum _{<xy>} \sigma _x \sigma _y \; - \; N \, 
\bar{s} \right)^2  \right\rangle \; = \; 
\left\langle \left[ \sum _{x} \left({\cal A}(x) \; - \; \bar{\cal A} \right)
\right]^2  \right\rangle 
\label{eq:mc4} \\
&=& \sum_{x,y} \, \left\langle \left( {\cal A}(x) \; - \; \bar{\cal A} 
\right)  \left( {\cal A}(y) \; - \; \bar{\cal A} \right) \right\rangle
\; .
\label{eq:mc5} 
\ena  
The crucial observation is that the connected correlation 
function 
\be 
D(x-y) \; := \; \Bigl\langle \left( {\cal A}(x) \; - \; \bar{\cal A} 
\right)  \left( {\cal A}(y) \; - \; \bar{\cal A} \right) \Bigr\rangle 
\label{eq:mc6} 
\en 
exponentially decreases for large values of $\vert x - y \vert $, i.e. 
$D(x) \propto \exp \{ - x/\xi _A \}$, where $\xi _A$ is the
correlation length characteristic for fluctuations in the action density. 
Hence, one finds that 
its integrated strength, the so-called {\tt susceptibility} $\rho $, is finite 
at least for $\beta \not= \beta _c$, i.e., 
\be 
\rho \; := \; \sum _x \; D(x) \; < \; \infty \; . 
\label{eq:mc7} 
\en 
These findings tell us that the standard deviation $\delta $ 
(\ref{eq:mc5}) linearly grows with the number of sites, i.e. 
$ \delta ^2\; = \; N \, \rho $. 

\vskip 0.3cm 
Using the central limit theorem to estimate the probability 
for accepting an action density $s$, we find 
 \be 
P_A \approx \exp \biggl( - \frac{ ( N s - N \bar{s} )^2 }{\delta^2 } 
\biggr) \; = \; 
\left[ \exp \left( - \frac{ ( s - \bar{s} )^2 }{\rho  } \right) 
\right] ^N \; . 
\label{eq:mc8} 
\en 
Hence, in the case of many sites, only configurations with an action 
per site close to the average action density  can significantly contribute 
to the partition function. If we randomly choose the spins on the sites the 
action density can take any value between $-1$ and $1$, and 
the argument $ (s - \bar{s} )/\rho$ is generically of order $1$. 
Hence the acceptance rate is down to $\mathrm{e}^{- 128 \times 128 } 
\approx 10^{-7115}$. 

\vskip 0.3cm 
The basic idea is to only generate configurations which are relevant. 
Starting from a seed configuration $c_0$, we will generate 
subsequent configurations $c_1$, $ c_2 $, \ldots , where 
the result for $c_{i+1}$ should only depend on the precessing 
configuration $c_i$ and must not depend on the configurations 
$c_{i-1}$. In this case, the set of configurations, 
$$ 
c_0  \; \longrightarrow \; c_1 \;  \longrightarrow \;
c_2  \; \longrightarrow \; c_3 \;  \longrightarrow \; \ldots \; 
 \longrightarrow \; c_\infty 
$$
is called a {\it Markov chain}. Central ingredient to a Markov chain 
is the probability $W(b,a)$ with which configuration $b$ is created out 
of configuration $a$. This probability must satisfy certain constraints: 

\bigskip
\begin{center}
\begin{tabular}{lll}
(i)   & Normalisation & $ \sum _b W(b,a) = 1, \; \forall a $ \\
(ii)  & Ergodicity    & $ W(b,a) > 1 , \;   \forall a,b $ \\ 
(iii) & Stability     & $ \sum _a W(b,a) \, P(a) = P(b), \,  \forall b$ ,  
\end{tabular} 
\end{center}

\bigskip
where $P(a)$ is given in (\ref{eq:mc1}). If these conditions are met, 
the series $c_i$ converges to a configuration which is distributed 
according to $P( c_\infty )$ (\ref{eq:mc1}). In order to see this, 
we introduce the probability $Q_i(c)$ for finding a configuration $c$ 
at position $i$ of the Markov chain, and denote the deviation from
the desired distribution by 
\be 
\epsilon _i \; = \; \sum _c \Bigl\vert Q_i(c) \, - \; P(c) \Bigr\vert \; . 
\label{eq:mc9} 
\en 
Because of property (ii), there is a $W_\mathrm{min}$ with 
\be 
W (a,b) \, \ge \, W_\mathrm{min} \, > \, 0 \; , \hbo 
W^\prime (a,b) := W (a,b)  - W_\mathrm{min} \, \ge \, 0 \; . 
\label{eq:mc11} 
\en 
Furthermore, the condition (i) implies that 
\be 
\sum _c Q_i(c) \; = \; 1 \; , \hbo \hbox{as} \; \; \; 
\sum _c P(c) \; = \; 1 \; . 
\label{eq:mc12} 
\en 
Using the stability condition (iii), we then obtain: 
\bea
\epsilon _{i+1} &=& \sum _c \Bigl\vert \sum _a W(c,a) Q_i(a) \, - \; 
P(c) \Bigl\vert \; = \; \sum _c \Bigl\vert \sum _a W(c,a) \, 
\Bigl[ Q_i(a) \, - \; P(a) \Bigr]  \Bigl\vert 
\nonumber \\ 
&=& \sum _c \Bigl\vert \sum _a W^\prime (c,a) \, 
\Bigr[ Q_i(a) \, - \, P(a) \Bigl] \; 
+ \;  W_\mathrm{min} \, \sum _a [ Q_i(a) \, - \; P(a) ] \Bigl\vert 
\nonumber \\ 
&=& 
\sum _c \Bigl\vert \sum _a W^\prime (c,a) \, 
\Bigr[ Q_i(a) \, - \, P(a) \Bigl] \; \Bigr\vert . 
\label{eq:mc15} 
\ena 
Using the triangle inequality and the positivity of $W^\prime $, we find 
\be  
\epsilon _{i+1} \; \le \; 
\sum _c \sum _a W^\prime (c,a) \, \Bigl\vert 
 Q_i(a) \, - \, P(a) \; \Bigr\vert . 
\label{eq:mc16} 
\en 
Changing the order of summation and using (see (\ref{eq:mc11})) 
\be 
\sum _c  W^\prime (c,a) \; = \; \sum _c 
( W (c,a) - W_\mathrm{min}) \, = \; 1 \, - \, 
n _\mathrm{conf} \, W_\mathrm{min} \; , 
\label{eq:mc17} 
\en 
where $n _\mathrm{conf}$ is the number of configurations, we finally find 
convergence: 
\be  
\epsilon _{i+1} \; \le \; [1 \, - \, n  _\mathrm{conf} 
\, W_\mathrm{min}] \; 
\sum _a \, \Bigl\vert  Q_i(a) \, - \, P(a) \; \Bigr\vert  \; = \; 
[1 \, - \, n _\mathrm{conf} \, W_\mathrm{min}] \; \epsilon _i \; . 
\label{eq:mc18} 
\en 
Instead of demanding the less stringent condition (iii), 
one often demands {\it   detailed balance}:
$$
\hbox{(iii)} \, ^\prime  \hbo 
W(b,a) \; P(a) \; = \; W(a,b) \; P(b) \; . 
$$
The latter condition immediately leads to condition (iii) if we sum 
the equation (iii)$\, ^\prime $ over the configurations $a$: 
$$ 
\sum _a W(b,a) \; P(a) \; = \; \sum _a W(a,b) \; P(b) \; = \; 
P(b) \; , 
$$
where we have used condition (i). Since condition (iii) follows from 
(iii)$\, ^\prime $ and only (iii) is necessary for our proof above, 
demanding  {\it   detailed balance}, i.e., (iii)$\, ^\prime$, is 
more restrictive.

\subsection{Heat-bath algorithm }

The heat-bath algorithm works as follows: (i) randomly choose 
a site $x_0$ and consider the corresponding spin $\sigma (x_0)$ 
for the update. Since the spin only interacts with its nearest 
neighbours, the interaction can be written as 
\bea 
H \; = \; \hbox{const.} \; - \; h_0 \, \sigma _{x_0} \; , 
\hbox to 1cm {\hfill } 
h_0 \; = \; \sum _{<x x_0>} \sigma _x \; . 
\label{kl:5} 
\ena 
The relative probability for choosing $\sigma (x_0) =1$ is given 
by $\exp \{ h_0 \beta \}$, and the relative probability for 
$\sigma (x_0) = -1 $ is given by $\exp \{ - h_0 \beta \}$. 
(ii) Calculate the absolute probability 
\bea 
p \; = \; \frac{ \exp \{ \beta h_0 \} }{ \exp \{ - \beta h_0 \} + 
\exp \{ \beta h_0 \} } 
\label{kl:6}
\ena 
with which the spin $\sigma _{x_0}$ must be set to $1$. 
Choose a random number $z \in [0,1]$. For $z<p$, choose $\sigma _{x_0}=1$ 
otherwise choose $\sigma _{x_0}=-1$.
(iii) Subsequently, pick another spin for the update and start again 
with (i). Once all spins have been visited at least once, 
one {\it sweep } has been performed. 

\vskip 0.3cm 
The algorithm above needs an initial configuration. 
We could choose a unique orientation of all spins. Since this 
is the ground state of the Hamiltonian which dominates the 
partition function for small values of the temperature, this initialisation  
is called a {\it cold start}. Alternatively, we could start with a 
random orientation of the spins. This is a configuration which is 
relevant at very high temperatures where interactions are negligible. 
This initialisation  is therefore called a {\it hot start}. 
Independently of our choice, a number of sweeps is carried out 
to generate a statistically relevant configuration. This procedure 
is known as {\it thermalisation}. The number required 
to arrive at an equilibrated spin lattice depends on the number 
of degrees of freedom and the temperature. 

\vskip 0.3cm 
\begin{figure}[t]
\centerline{ 
\epsfxsize=.7\textwidth
\epsffile{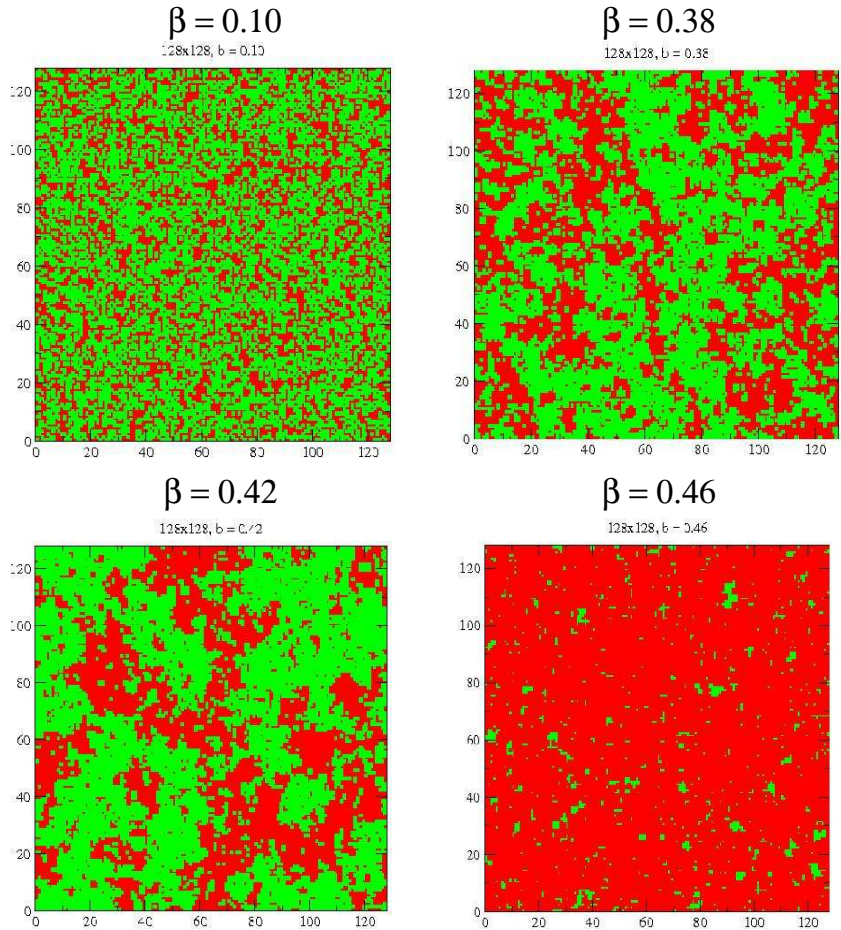}
}
\caption{ Thermalised spin configurations of the 2d Ising model for several 
$\beta $ values starting from high temperature phase to the 
low temperature ordered phase. 
}
\label{fig:is_mag} 
\end{figure}
Let us now examine typical lattice spin configurations. 
Starting at low $\beta $, the sample configurations are highly 
disordered (see figure~\ref{fig:is_mag}). 
Increasing $\beta $ up to $\approx 0.3$, the clusters 
of spins with the same orientation already extend over several 
lattice spacings. 
Approaching the critical value, e.g.~for 
$\beta \approx 0.42$, the clusters are already as large as the 
lattice. This observation reflects the growth of the spin 
correlation length which, for the present case, is  
\be 
\xi \; \approx \; \xi_+ \, \left\vert 1 - \frac{T}{T_c} \right\vert ^{- 1 } 
\; , \hbox to 1cm {\hfill } 
(T \stackrel{>}{_\sim}T_c) \; , 
\label{eq:mc20} 
\en
and hence diverges when $\beta \to \beta _c$. 

\vskip 0.3cm 
If our numerical approach should produce the 
configurations of a Markov chain, the configurations may not depend 
on the Monte-Carlo history. To find out whether the configurations are indeed 
statistically independent, we may inspect the autocorrelation time $\tau $ 
e.g.~say for the magnetisation $M$ (recall subsection~\ref{sec:auto} for 
discussions of autocorrelations). To guarantee independence, 
we perform about $2\tau $ Monte-Carlo sweeps before we consider 
a configuration eligible for contributing to an estimator. 
In the case of the heat bath algorithm (in fact, for all 
local update algorithms), one discovers that the autocorrelation time 
strongly increases when the critical point is approached. 
This implies that the interesting regime of the model, namely the 
regime close to the phase transition, is not accessible with 
these types of algorithm. The reason for this is the following: 
consider a spin inside one of the clusters. All the neighbouring spins are 
pointing in the same direction. If this spin is now subjected to a 
local update procedure, the spin hardly changes because of the 
strong mean field produced by the other spins. Hence, only the 
boundaries of the cluster are significantly modified after one sweep 
through the lattice. The correct physics is, however, described by 
configurations consisting of strongly fluctuating clusters. 
In order to change a cluster completely, there are roughly 
$\xi ^2$ lattice sweeps necessary. Hence, only after $\xi ^2$ sweeps, 
the configuration has changed significantly. This, however, implies 
that the autocorrelation time is roughly given by 
$\tau \approx \xi ^2$. Indeed, it was empirically observed for the Metropolis 
algorithm that 
\be 
\tau \; \approx  \; \xi ^z \; , \hbo z_\mathrm{Metro} 
\approx 2.125 \; . 
\label{eq:mc21} 
\en
The index $z$ is called the {\it dynamical critical exponent} and 
depends on the algorithm. Since the physical correlation length 
$\xi $ diverges at the phase transition, (\ref{eq:mc21}) 
implies that the regime near the phase transition cannot be simulated 
with local update algorithms. 

\subsection{Cluster update algorithms }

State-of-the-art simulations which explore the transition regime 
use the so-called cluster algorithms. The difference to 
local update algorithms is that many spins are flipped at a time. 
To derive the prescription of such a cluster update, we rewrite 
the partition function (\ref{kl:3}) as 
\bea 
{\cal Z} = \sum_{\{ \sigma _x \}} {\exp\left( \beta \, 
\sum_{<xy>} \sigma _x \sigma _y \right)} \; = \; 
\sum_{\{ \sigma _x \}} \prod _{<xy>} {\exp\bigl( \beta \, 
\sigma _x \sigma _y \bigr)} \; . 
\label{kl:10}
\ena 
If both neighbouring spins, $\sigma _x $ and $\sigma _y$, in (\ref{kl:10}) 
are equal, the probabilistic factor in (\ref{kl:10}) equals $\exp \beta $. 
For an opposite orientation of the spins,  the probabilistic factor 
is given by $\exp (- \beta )$. We therefore cast (\ref{kl:10}) into 
\bea 
{\cal Z} = \sum_{\{ \sigma _x \}} \prod _{<xy>} e^\beta \; 
\biggl[ (1 - p) \, + \, p \, \delta _{\sigma _x \sigma _y } \biggr] \; , 
\hbox to 1cm {\hfill } 
p := 1 \, - \, e^{-2 \beta } \; . 
\label{kl:11}
\ena 
We now are going to make the representation of the partition function 
even more involved by using the identity 
\bea 
a \, + \, b \; = \; \sum _{n=0}^{1} \biggl[ a \, \delta _{n0} \; + \; 
b \, \delta _{n1} \biggr] \; . 
\nonumber 
\ena
Introducing variables $n_{xy} \in \{0,1\}$, which 
are associated with each link of the lattice, we obtain 
\bea 
{\cal Z} = \sum_{\{ \sigma _x \}} \sum _{\{n_{xy}\} } 
\prod _{<xy>} e^\beta \; \biggl[ 
(1-p) \, \delta _{n_{xy} , 0 } \; + \; p \, \delta _{\sigma _x \sigma _y} 
\, \delta _{n_{xy},1} \biggr] \; .
\label{kl:12}
\ena 
The cluster update prescription is now obtained by performing 
standard heat bath steps for the variables  $\{ \sigma _x \}$ and 
$\{n_{xy}\}$.

\vskip 0.3cm 
Let us consider the update for the link variables $n_{xy}$ first. 
In order to avoid generating a configuration of vanishing 
probability, we must choose $n_{xy} = 0$ if the neighbouring 
spins $\{ \sigma _x \}$ and $\{ \sigma _y \}$ are different (see 
(\ref{kl:12})). If these spins are equally oriented, the probabilistic 
measure in (\ref{kl:12})) is given by 
$$ 
(1-p) \, \delta _{n_{xy} , 0 } \; + \; p \, \, \delta _{n_{xy},1} \; , 
$$
implying that the link $n_{xy}$ is set to $1$ with probability $p$.
Given an initial spin distribution, the values of all link variables 
can be chosen according to the above prescription. 

\vskip 0.3cm 
Let us now consider the spin update. According to probabilistic 
measure, i.e., 
$$ 
(1-p) \, \delta _{n_{xy} , 0 } \; + \; p \, \delta _{\sigma _x \sigma _y} 
\, \delta _{n_{xy},1} \; , 
$$
all spins which are connected by links $n_{xy}=1$ must be of equal 
orientation. All spins which are connected by so-called activated links, 
i.e.,  $n_{xy}=1$, 
are said to be part of a cluster. The task is now to find all such clusters 
on the lattice. Once these clusters have been identified, we assign 
$\pm 1$ (with equal probability) to all spins of the same cluster. 
 
\vskip 0.3cm 
This first versions of such cluster update algorithms are due to 
Fortuin and Kasteleyn~\cite{Fortuin:1971dw}, Swendsen and 
Wang~\cite{Swendsen:1987ce} and Wolff~\cite{Wolff:1988uh}. 
It is found empirically that the dynamical critical exponent is 
strongly reduced: 
\be 
\tau \; \approx \; \xi ^z \; , \hbo 
z_\mathrm{cluster} \approx 0.2 \; . 
\label{kl:13} 
\en 
Introductory discussions can be found 
in~\cite{Fortuin:1971dw,Swendsen:1987ce,Wolff:1988uh,Janke}.

\section{Quantum field theories on computers}

\subsection{Quantum mechanics } 
\label{sec:1.1}

Let us assume that the motion of a particle of mass $m$ in 1 dimension 
is governed by a potential $V(x)$. The classical equation of motion 
can be calculated by variational methods from the action 
\be 
S \; = \; \int _0^t dt \; \Bigl\{ \frac{m}{2} \dot{x}^2 \; - \; V(x) 
\Bigr\} \; . 
\label{eq:1} 
\en 
Classically, these equations of motion determine the time evolution 
of the position of the particle $x(t)$. 
At quantum mechanical level, the partition function 
\be 
Z(T) \; = \;  \tr \; \exp \biggl\{ - \frac{1}{T} 
\, H \, \biggr\}  
\label{eq:2} 
\en
is a convenient starting point to discuss the thermodynamics 
of the physical system. Here, $H$ is the quantum mechanical 
Hamiltonian, i.e., 
\be 
H \; = \; - \frac{\hbar ^2 }{2m} \frac{d^2 }{dx^2} \; + \; V(x) \; . 
\label{eq:3} 
\en 
$T$ is the temperature, and is considered as an external parameter. 
Once one has succeeded in calculating the partition function (\ref{eq:2}), 
thermodynamical quantities can be easily obtained by taking 
derivatives, e.g., the temperature dependence of the internal energy 
is given by 
\be 
\langle H \rangle \; = \;  T^2 \, \frac{ d \, \ln Z(T) }{ dT } \; . 
\label{eq:4} 
\en 
Although a direct calculation of the eigenstates $\langle n \vert $ 
of the Hamiltonian 
might be the easiest way to calculate a quantum mechanical partition 
function in practical applications, I would like to reformulate (\ref{eq:2}) 
in terms of a functional integral. This will be of great help when we 
generalise the quantum mechanical considerations to the case of a 
quantum field theory. 

\vskip 0.3cm 
For these purposes, I introduce a length scale $L:=1/T$ and an interval 
$[0,L]$ which I decompose into $N$ equidistant portions of length 
$a \ll L$, where $a$ is called {\it lattice spacing}. It is trivial to obtain 
\be 
\exp \biggl\{ - \frac{1}{T} \, H \, \biggr\}  \; = \; 
\exp \biggl\{ - \sum_{\nu =1}^N \, a \, \, H \, \biggr\} 
\; = \; \prod _{\nu=1}^N \; \exp \{ - \,a \, H \} \; . 
\label{eq:5} 
\en  
Let us define complete sets of momentum and position eigenstates 
($\vert p\rangle $ and $\vert x \rangle $, respectively) by 
\be 
1 \; = \; \int dx_\nu \; \vert x_\nu \rangle \, \langle x_\nu \vert \; , 
\hbo 
1 \; = \; \int dp_\nu \; \vert p_\nu \rangle \, \langle p_\nu \vert \; , 
\en 
for $\nu = 1 \ldots N $. As usual, these states obey 
$$
\langle p_k \vert x_k \rangle = \exp \Bigl\{ - \frac{i}{\hbar } p_k x_k 
\Bigr\} \; . 
$$ 
Using a complete set $\vert x_0 \rangle $ of position  eigenstates 
to evaluate the trace in (\ref{eq:2}), we find 
\bea
\int dx_0 && \langle x_0 \vert \; \prod _{\nu=1}^N \exp \{  - aH \} 
\vert x_0 \rangle \; = \; \int dx_0 \, dp_0 \; dx_1 \; dp_1 \; 
\ldots \; dx_{N-1} \, dp_{N-1} \; 
\nonumber \\ && 
\langle x_0 \vert \mathrm{e}^{  - aH } \vert p_0 \rangle \, 
\langle p_0 \vert x_1 \rangle \; 
\langle x_1 \vert \mathrm{e}^{  - aH } \vert p_1 \rangle \, 
\langle p_1 \vert x_2 \rangle \; \ldots \; 
\langle x_{N-1} \vert \mathrm{e}^{  - aH } \vert p_{N-1} \rangle \, 
\langle p_{N-1} \vert x_0 \rangle \; . 
\nonumber
\ena 
Note that the operators $p^2$ and $V(x)$ do not commute. We may, however, 
write: 
\be 
\exp \left\{ - a \frac{p^2}{2m} - a V(x) \, + \, 
\frac{a^2}{4m} [V(x), p^2] \, + \, \ldots \right\} 
\; = \; 
\exp \left\{  - a V(x)  \right\} \; 
\exp \left\{   - a \frac{p^2}{2m} \right\} \; . 
\nonumber 
\en 
Since $\vert  x \rangle $ and $\vert p \rangle $ are eigenstates of the 
position operator and momentum operators, respectively, 
we find 
\be 
\langle x_{k} \vert  \exp \{  - aH \} \vert p_{k} \rangle 
\; = \; \exp \left\{ -a \left[ \frac{p_k^2}{2m} \, + \, V(x_{k}) 
\, + \, {\cal O}(a) 
\right] \right\} \; \exp \{\frac{i}{\hbar } p_{k} x_{k} \}  \; . 
\nonumber 
\en  
The partition function therefore becomes up to terms of order $a^2$
\bea 
Z(T) &=& \int dx_0 \, dp_0 \; dx_1 \; dp_1 \; 
\ldots \; dx_{N-1} \, dp_{N-1} \; dx_N
\exp \left\{ -a \sum _{k=0}^{N-1} \left[ \frac{p_k^2}{2m} \, + \, V(x_{k}) 
\right] \right\} 
\nonumber \\ 
&& \exp \left\{
- \frac{i}{\hbar } \sum _{k=0}^{N-1} p_{k} (x_{k+1} - x_{k}) \right\} \; 
\langle x_0 \vert x_N \rangle 
\label{eq:6}
\ena
It is straightforward to perform the momentum integrations, which 
are Gaussian,  
\bea 
Z(T) &=& \left( \frac{ 4 \pi m }{a} \right)^{N/2} 
\int dx_0 \; dx_1 \ldots dx_N \; \delta _{x_0 x_N} 
\label{eq:7} \\
&& \exp \left\{ -a \sum _{k=0}^{N-1} \left[ \frac{m}{2} \frac{(x_{k+1}
- x_k)^2}
{a^2 \, \hbar ^2 } \, + \, V(x_{k}) \right] \right\} 
\nonumber 
\ena
This equation is a completely regularised expression for 
the partition function and can be directly used 
in numerical simulations. Note that in the framework of quantum field 
theory, one sets $\hbar =1$. 

\vskip 0.3cm 
A compact notation can be derived by formally taking the 
lattice spacing $a$ to zero. For this purpose, we define $a_h := \hbar 
\, a$, and the Euclidean action by 
\be 
S_E \; = \; \int _0^L d\tau \biggl\{ \frac{m}{2} \dot{x}^2 \; + \; 
V(x) \biggr\} \; . 
\label{eq:8} 
\en 
Note the sign change in front of the potential compared with 
the standard action (\ref{eq:1}). The interval $[0,L]$, which was 
introduced above (\ref{eq:5}), is called {\it Euclidean time } interval. 
By construction (see above), the length of the Euclidean time interval is 
given by the inverse temperature, i.e., $L=1/T$. We also introduce 
a Euclidean particle trajectory, and a Euclidean velocity 
\be 
x_k \rightarrow x(\tau ) \; \hbo 
\frac{ x_{k+1} - x_k }{a_h} \rightarrow  \dot{x}(\tau ) \; , 
\en 
where we identify $d\tau =a_h$. Using the shorthand notation 
$$ 
\left( \frac{ 4 \pi \hbar \, m }{a_h} \right)^{N/2} 
\int dx_0 \; dx_1 \ldots dx_{N-1} \; \rightarrow \; {\cal D} x(\tau ) \; , 
$$ 
the partition function (\ref{eq:7}) can be formally written as a functional 
integral 
\be 
Z(T) \; = \; \int {\cal D} x(\tau ) \; 
\exp \biggl\{ - \frac{1}{\hbar } \, S_E \biggr\} \; . 
\label{eq:9} 
\en 
Eq.(\ref{eq:9}) suggests that an average over all Euclidean 
trajectories $x(\tau )$ must be performed where the probabilistic weight 
of each trajectory is given by $\exp \{ - S_E /\hbar \}$. Note also that 
in view of the $\delta $-function in (\ref{eq:7}) only 
trajectories which are periodic in Euclidean time must be considered, i.e., 
$x(0) = x(L=1/T)$. 

\subsection{Quantum field theory }

\begin{figure}[t]
\centerline{ 
\epsfxsize=12cm
\epsffile{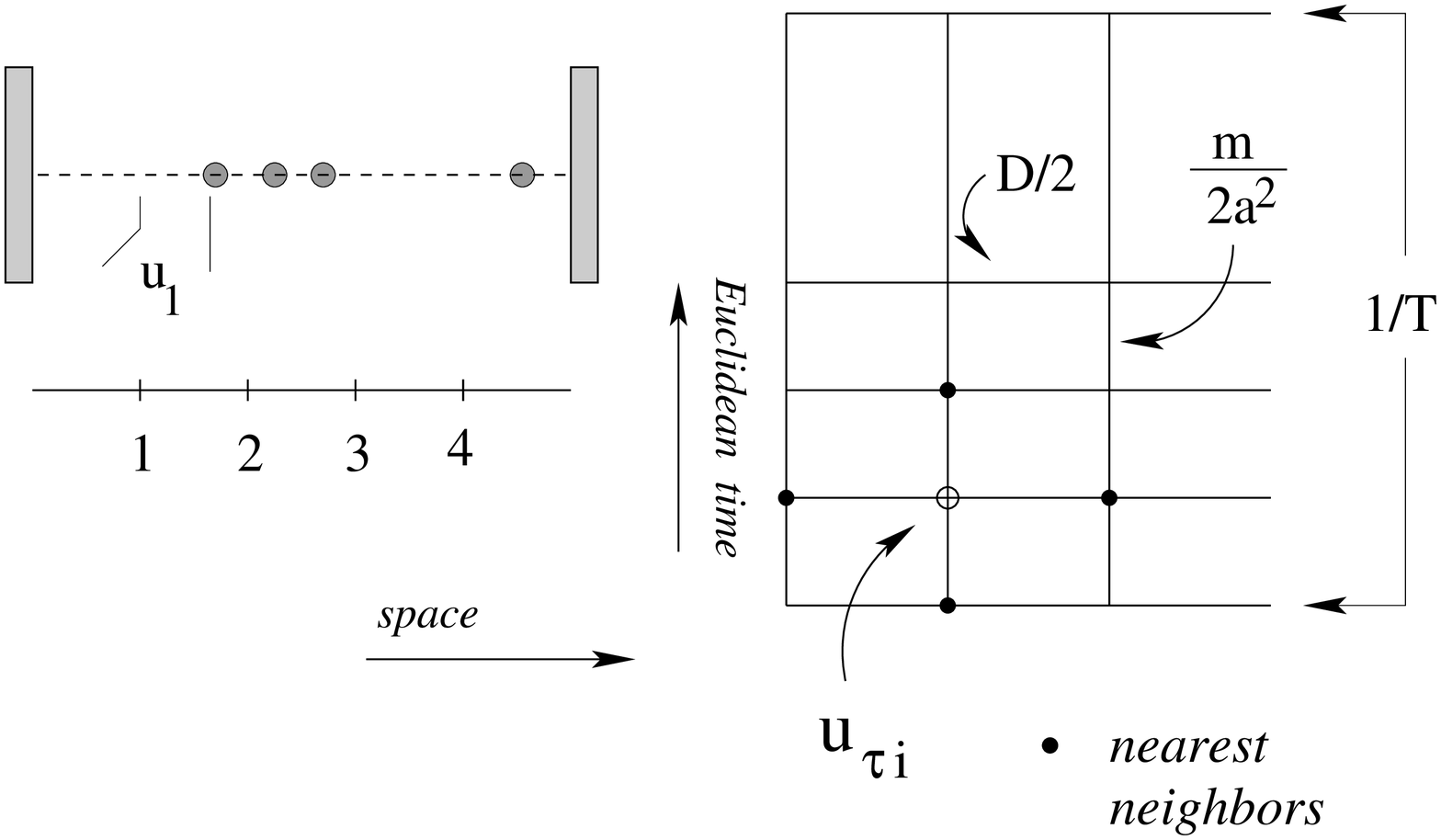}
}
\caption{ Classical versus quantum partition functions of a 1-dimensional 
   particle chain. } 
\label{fig:2} 
\end{figure}

\vskip 0.3cm 
For illustration purposes, we consider the 1-dimensional 
particle chain in figure~\ref{fig:2}. 
Here, the positions of the particles $i = 0 \ldots n$ are 
characterised by their extensions  $u_i $ from the equilibrium 
position. The particles experience a harmonic potential depending on 
the distance to the nearest neighbour. Here, I choose the boundary 
conditions $u_0=0$, $u_n=0$. The Hamiltonian, which describes 
the classical physics, is given by 
\be 
{\cal H} \; = \; \sum _{i =1 }^{n-1} \left[ \frac{1}{2m} p_i ^2 
\; + \; \frac{D}{2} \left( u_{i +1 } - u_i \right)^2 \right] \; .
\label{eq:20} 
\en 
Hence the classical partition function is given by the multi-dimensional 
integral 
\be 
Z_{cla}(T) \; \propto \; \int dp_1 \ldots dp_{n-1} \; du_1 \ldots 
du_{n-1} \; \exp \biggl\{ - \frac{ \cal H}{T} \biggr\} \; . 
\label{eq:21} 
\en 
In order to calculate the full quantum mechanical partition function 
of the particle chain, we first write down the Euclidean partition 
function. Note for this purpose that the displacements $u_i$ now 
acquire an additional dependence on the Euclidean time 
$u_i \rightarrow u_i(\tau ) \equiv u_{\tau i}$. With this notation the 
Euclidean action is given by 
\be 
S_E \; = \; \sum _{\tau =1}^N \sum _{i=1}^{n-1} a \left[ 
\frac{ m }{2a^2 } \left( u_{\tau, \, i } - u _{\tau -1, \, i} \right)^2 
\; + \; 
\frac{ D }{2 } \left( u_{\tau , \,  i+1 } - u _{\tau , \, i} \right)^2 
\right] \; . 
\en 
The interactions between the c-number fields $u_{\tau i}$ can be 
easily visualised (see figure \ref{fig:2}): the fields $u_{\tau i}$ 
harmonically interact with their nearest neighbours. The harmonic 
interaction strength is given by $D/2$ in space direction and 
$m/2a^2$ for neighbours in Euclidean time direction. 
The quantum mechanical partition function can be calculated by 
integrating over of the fields $u_{\tau i}$ located at the sites 
of a 2-dimensional grid, .i.e. 
\be 
Z(T) \; \propto \; \int {\cal D} u \; \exp \{ - S_E \} \; , 
\label{eq:22} 
\en 
where the temperature enters the consideration via the extension 
of the lattice in Euclidean time direction with fields obeying 
periodic boundary conditions. 

\vskip 0.3cm 
To conclude, we observe that the partition function of a classical 
$D+1$ dimensional field theory (in lattice regularisation) 
describes the full partition function of a $D$ dimensional quantum 
system. $D$ is the number of space dimensions. This correspondence 
is very helpful in understanding the quantum behaviour of a theory, since 
it can be mapped to a classical field theory (at the expense of an 
additional dimension). In the next section, we will study 
classical partition functions in 4-dimensional Euclidean space 
in order to derive the information on the thermodynamics of the 
full quantum system. 

\section{Lattice gauge theory}

\subsection{The gauged Ising model}

The Ising model, strictly speaking the partition function (\ref{kl:3}), 
is invariant under the {\it global} transformation of the spins given by 
\be 
\sigma ^\Omega (x) \; = \; \Omega \; \sigma (x) \; , \hbo 
\Omega \; = \; \pm 1 \; . 
\label{eq:gi1}
\en 
The transformation is called {\it global} because the transformation 
affects all spins at the same time, i.e., $\Omega $ is independent 
of the coordinates (sites). The corresponding symmetry group is $Z_2$. 

\vskip 0.3cm 
This symmetry group can be generalised to a {\it local} symmetry, also 
known as {\it gauge symmetry}, by demanding invariance under 
\be 
\sigma ^\Omega (x) \; = \; \Omega(x) \; \sigma (x) \; , \hbo 
\Omega (x) \; = \; \pm 1 \; . 
\label{eq:gi2}
\en 
Of course, the action (\ref{kl:4}) of the standard Ising model 
is not invariant under the huge symmetry group which is now 
$[Z_2]^N$, where $N$ is the number of sites. In order to obtain 
a version of the Ising model which possesses a $Z_2$ gauge symmetry, 
we need to change the action. The only way to do it, is to introduce 
an additional field, $Z_\mu (x) $. This field is associated with 
the links of the lattice: $x$ specifies the site and $\mu $ the direction 
in which we find the link. Alternatively, we could write: 
$$ 
Z_\mu (x) \; = \; Z_{\langle xy \rangle} \; , \hbo y \; = \; x + \hat{e}_\mu 
\; , 
$$
where $\hat{e}_\mu$ is the unit vector in $\mu $ direction. For 
the latter expression, we will also abbreviate 
$$ 
 x + \hat{e}_\mu  \; = \; x \, + \, \mu \; . 
$$
For the action, we choose 
\be 
S_\mathrm{matter} 
\; = \; \kappa \, \sum _{\langle xy \rangle } \sigma (x) \, Z_\mu (x) \, 
\sigma (x+\mu) \; , 
\label{eq:gi3}
\en 
and demand that the link $Z_\mu $ transforms under gauge transformations 
as 
\be 
Z^\Omega _\mu (x) \; = \; \Omega (x) \; Z_\mu (x) \; \Omega (x+\mu ) \; . 
\label{eq:gi4}
\en 
Since spin and link transform simultaneously with the same $\Omega (x)$ and 
since $\Omega ^2 (x) = 1$, one easily proves the gauge invariance of the 
action (\ref{eq:gi3}). 

\vskip 0.3cm 
Obviously, the action $S_\mathrm{matter} $ describes the interaction 
between the {\it matter} fields, i.e., the spins, and the new link 
fields. What is left to do is to design a gauge invariant action for these
new degrees of freedom. This interaction should be short ranged in 
order to preserve some desirable features such as universality. 
A possible choice is 
\be 
S_\mathrm{link} \; = \; \beta 
\sum _{x, \, \mu > \nu } P_{\mu \nu }(x) \; , \hbo 
P_{\mu \nu }(x) \; = \; Z_\mu (x) \, Z_\nu (x+\mu) \, 
Z_\mu (x+\nu) \, Z_\nu (x) \; . 
\label{eq:gi5}
\en 
Here, the numbers $(x, \, \mu > \nu)$ specify the plaquette of the lattice 
the lower left corner of which is located at site $x$ and which is spanned 
by the directions $\mu $ and $\nu $. The field combination $P_{\mu \nu }(x)$ 
is often called the plaquette for short. The proof that 
$P_{\mu \nu }(x)$ is indeed invariant under gauge transformations (\ref{eq:gi4}) 
is left to the reader. 

\vskip 0.3cm 
The total action of the gauged Ising model consists of two parts: the matter 
part and the ``gauge'' part. Correspondingly, there are two coupling 
constants: the convention is that $\beta $ is the pre-factor in the 
pure gauge action, while $\kappa $ multiplies the matter part. 

\vskip 0.3cm 
Once our system is now gauged, it only makes sense to consider 
gauge invariant observables since non-gauge invariant quantities 
vanish. Let us explore this for a simple gauge variant quantity such 
as the spin correlation function:
\bea 
C(x_0,y_0) &=& 
\frac{1}{ \cal Z } \; \sum _{ \{\sigma \} } \; \sigma (x_0) \, 
\sigma (y_0) \; 
\exp \Big\{ S[\sigma ] \Bigr\} \; , \hbo 
S[\sigma ] \; = \; S_\mathrm{matter} + S_\mathrm{link} \; , 
\label{eq:gi6} \\ 
 \cal Z  &=& \sum _{ \{\sigma \} } \; \exp \Big\{ S[\sigma ] \Bigr\} \; , 
\label{eq:gi7} 
\ena 
Let us now consider a particular gauge transformation (\ref{eq:gi2}) 
of the spins, i.e., 
\be 
\Omega (x) \; = \; \left\{ \begin{array}{l} 
-1 \hbo \hbox{for} \; \; x = x_0 \\ 
\phantom{-}1 \hbo \hbox{else} \; \end{array} \right. 
\label{eq:gi8} 
\en
Renaming all spins in the sum in (\ref{eq:gi6}) by 
$\sigma (x) \rightarrow \sigma ^\Omega (x) $, we use the gauge 
invariance of the action and the sum , i.e., 
$$ 
S[\sigma ] \; = \; S[\sigma ^\Omega ] \; , \hbo 
\sum _{ \{\sigma \} } \; = \; \sum _{ \{\sigma ^\Omega \} } \; . 
$$
The sum is trivially invariant, since we sum anyhow over all possible 
$\pm 1$ combinations for the spins. Thus, we obtain:
\bea
C(x_0,y_0) &=&
\frac{1}{ \cal Z } \; \sum _{ \{\sigma \} } \; \sigma ^\Omega (x_0) \, 
\sigma ^\Omega (y_0) \; \exp \Big\{ S[\sigma ] \Bigr\} 
\nonumber \\ 
&=& \frac{1}{ \cal Z } \; \sum _{ \{\sigma \} } \; [ - \, \sigma (x_0)]  \, 
\sigma (y_0) \; \exp \Big\{ S[\sigma ] \Bigr\} \; = \; 
- \, C(x_0,y_0) \; , 
\label{eq:gi10} 
\ena 
where we used our particular choice (\ref{eq:gi8}) in the last line above. 
We conclude from this that $C(x_0,y_0)=0$.

\subsection{Pure $Z_2$ gauge theory: 3 and 4 dimensions}
\vskip 0.3cm 
\begin{figure}[t]
\centerline{ 
\epsfxsize= 0.5 \textwidth
\epsffile{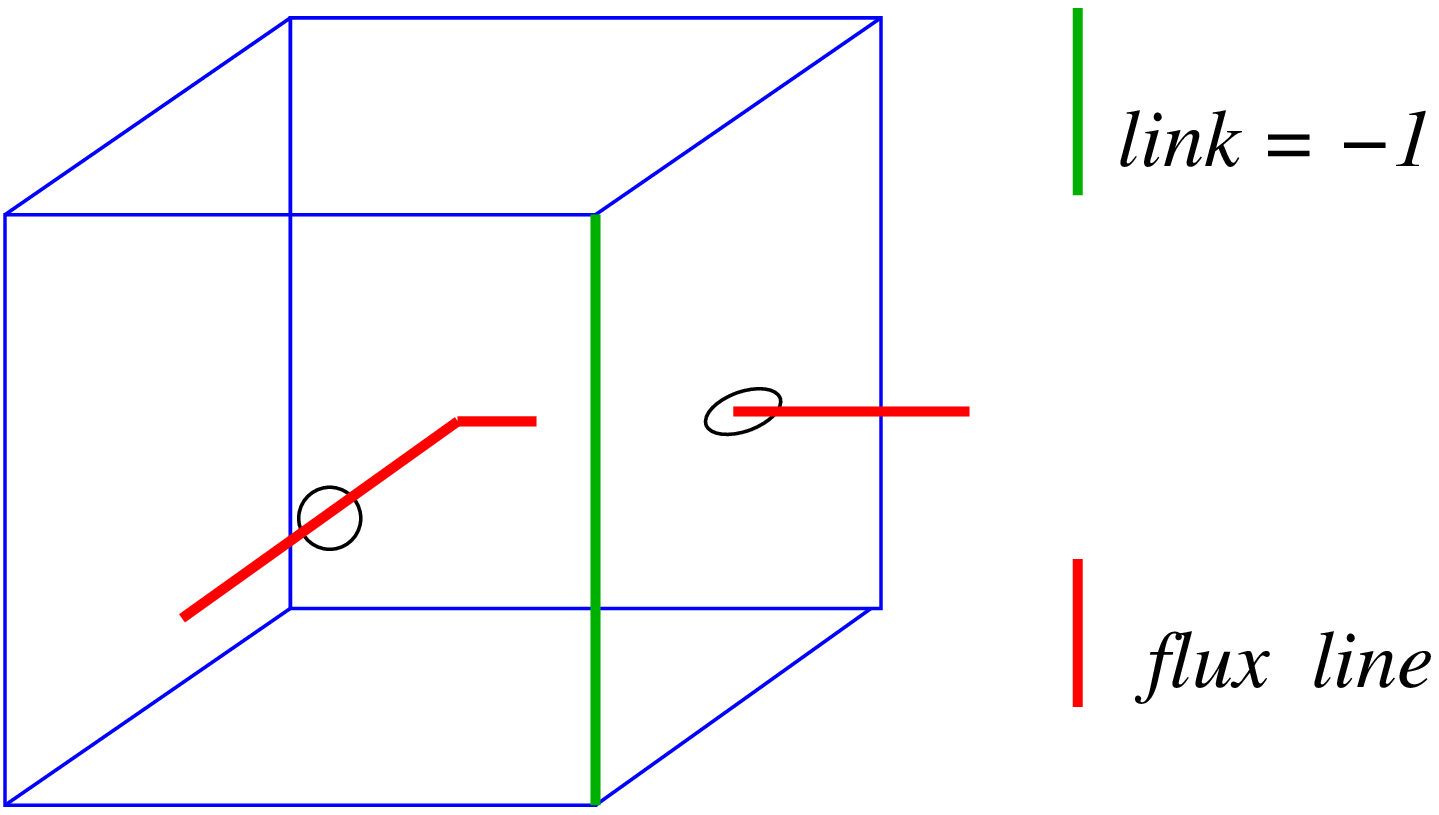}
\epsfxsize= 0.5 \textwidth
\epsffile{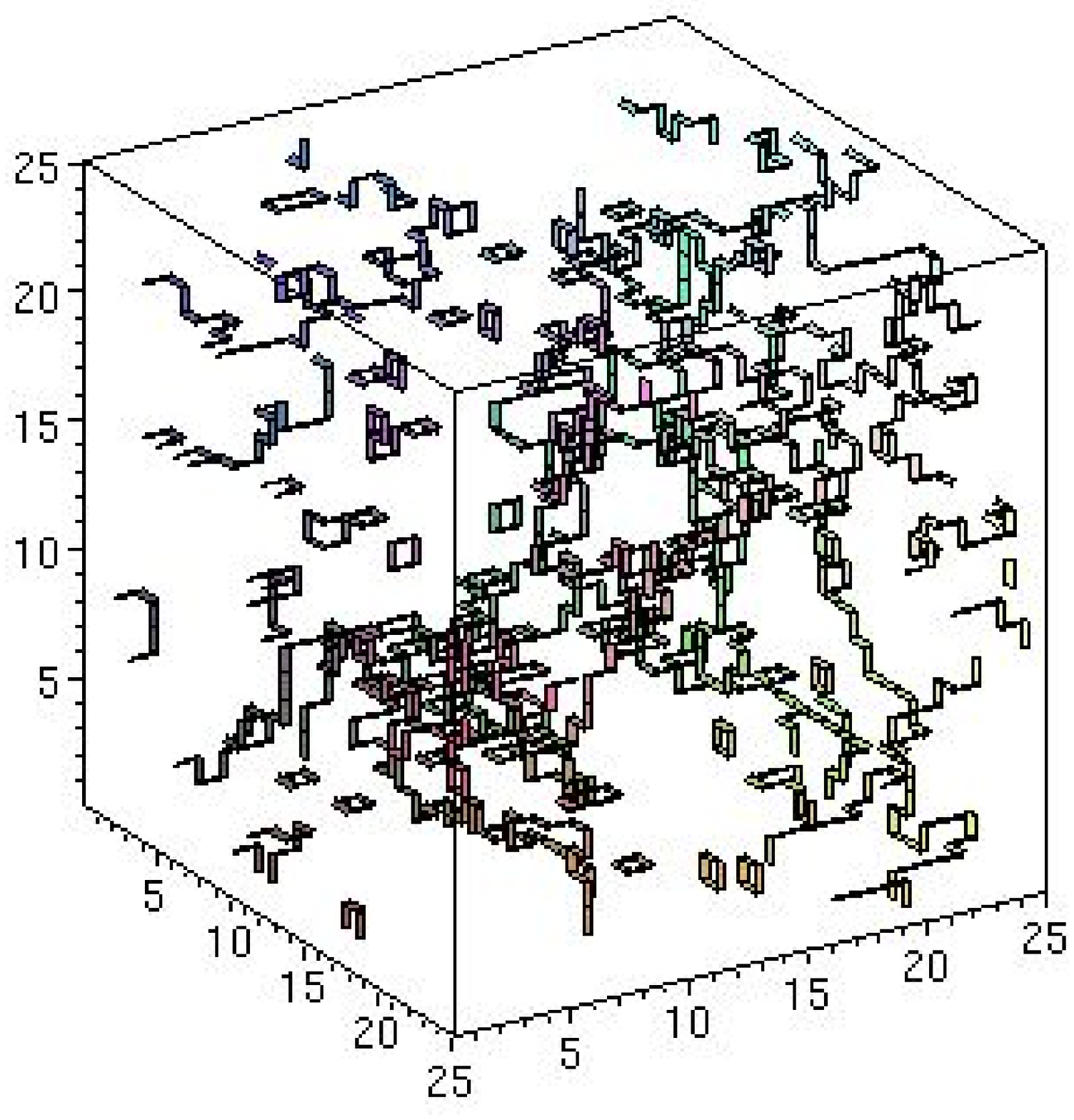}  
}
\caption{ Flux passing through an elementary cube (left) and 
closed flux lines on a 3d lattice (right).}
\label{fig:gi1} 
\end{figure}
Let us now consider the particular case $\kappa =0$ when the matter 
fields are absent from the theory. The emerging theory is 
called {\it pure $Z_2$} gauge theory, and it is a theory 
of link fields only. What are the physical (i.e.~gauge invariant) 
degrees of freedom in this case? Let us consider the more interesting case 
of 3 and 4 dimensions for these considerations. 
In order to talk about gauge invariant information, we now consider 
the plaquettes $P_p$, $p =(x; \mu \nu )$, defined in (\ref{eq:gi5}). 
We say that a short flux line (vortex) passes through the plaquette 
$p$ if $P_p = -1$. Since the plaquette variables $P$ are gauge invariant, 
so are the flux lines. More formally, we introduce a vortex 
plaquette variable by 
\be 
v_p \; = \; \prod _{l\in p} Z_l 
\label{eq:gi11} 
\en
and consider the flux lines which enter/leave an elementary 
cube of the lattice. We take the product of all vortex plaquettes which 
are associated with the faces of the elementary cube and find 
\be 
\prod _{p\in c} v_p \; = \; (-1)^\nu \; , 
\label{eq:gi12} 
\en
where $\nu $ is the total number of vortices at the faces of the cube. 
Inserting the definition (\ref{eq:gi11}), we also find that 
\be 
\prod _{p\in c} v_p \; = \; \prod _{p\in c} \prod _{l\in p} Z_l \; = 
\; 1 \; , 
\label{eq:gi15} 
\en
since in the latter products all $Z_l$ factors appear twice (see 
figure~\ref{fig:gi1}, left panel; remember $Z_l^2 =1$). 
Comparing (\ref{eq:gi15}) 
with (\ref{eq:gi12}), we realise that $\nu $ must be even. In particular, 
$\nu =1$ is excluded implying that a vortex never ends inside a cube. 
Considering 3 dimensions (or the spatial hypercube of 4-dimensional 
space time), we find that the gauge invariant vortices form closed lines 
in space. See figure~\ref{fig:gi1}, right panel, for an illustration. 

\vskip 0.3cm 
{\tt 4 dimensions: } Let us consider the 4 dimensional model first. 
The constraint (\ref{eq:gi15}) is most easily interpreted on the dual 
lattice. A plaquette $p$ of the original lattice maps onto a plaquette 
$^\ast p$ of the dual lattice, and a cube $c$ corresponds to a link 
$^\ast l$ on the dual lattice. Hence, the constraint  (\ref{eq:gi15}) 
reads on the dual lattice 
\be 
\prod _{^\ast p \in ^\ast l} v_{^\ast p} \; = \; 1 \; . 
\label{eq:gi16} 
\en
This simply means that the number of vortex plaquettes 
which are attached to a link on the dual lattice must be even. 
Accordingly, the vortex plaquettes form closed surfaces 
on the dual lattice. If $n$ denotes the number of negative plaquettes 
on the original lattice, the number of trivial plaquettes in 4 dimension 
is $6N -n$, where $N$ is the number of lattice points. Hence, 
the probabilistic weight of such a configuration is: 
$$ 
\left[ \exp \left\{ \beta \right\} \right] ^{6N - n} \; 
\left[ \exp \left\{ - \beta \right\} \right] ^{n} \; = \; 
\exp \left\{ 6N \, \beta \right\} \; \exp \left\{ -2  \beta \, n\right\} \; , 
$$
so that the partition function can be written as 
\be 
{\cal Z} \; = \; \exp \left\{ 6N\, \beta \right\} \; \sum _{ \{\hbox{closed 
 surfaces} \} } \;  \exp \left\{ -2  \beta \, n\right\} \; . 
\label{eq:gi17} 
\en 
Let us interpret this partition function: the degrees of freedom 
are closed two 
dimensional sheets (2-branes) embedded in four dimensions. 
The surface $A$ of these 
branes is given by $n$. Hence, the probabilistic factor is given by 
$$ 
\exp \Bigl\{ -2  \beta \, A \Bigr\} 
$$
implying that $ 2 \beta $ can be interpreted as the {\it surface tension}. 
At zero temperature $(\beta \to \infty)$, the empty vacuum 
(no 2-branes) is realised. At finite temperatures, the brane entropy 
competes with the penalty from the weight factor. A direct 
calculation of the entropy of 2d world-sheets in 4d would be cumbersome. 
However, exploiting the relation to the $Z_2$ gauge theory 
makes the calculation 
of brane expectation values easily accessible by numerical means. 

\vskip 0.3cm
Let us proceed to obtain the duality map of the 4d $Z_2$ gauge theory. 
The basic trick to perform the sum over the closed surfaces is to 
introduce degrees of freedom which automatically resolve the constraint. 
In the present case, these are links $^\ast Z_{\ast l}$ on the dual lattice.  
Let us consider 
\be 
\sum _{ \{ ^\ast Z_{\ast l} \} } \; \prod _{\ast p} 
\Bigl[ 1 \; + \; t \; P _{ \ast p} [^\ast Z ] \Bigr] \; , 
\label{eq:gi18} 
\en 
where $P _{ \ast p} [^\ast Z ] $ is the plaquette generated by the links 
$^\ast Z$ on the dual lattice. When we remove the brackets in (\ref{eq:gi18}), 
the only way to have a non-vanishing contribution to the sum is 
by making sure that each link $^\ast Z_{\ast l}$ appears an even number 
of times. This, however, means that the negative plaquettes 
$P _{ \ast p} [^\ast Z ]$ form closed surfaces. Hence, we find 
\be 
\sum _{ \{ ^\ast Z_{\ast l} \} } \; \prod _{\ast p} 
\Bigl[ 1 \; + \; t \; P _{ \ast p} [^\ast Z ] \Bigr] \; = \; 
2^{4N} \; \sum _{ \{\hbox{closed surfaces} \} } \; 
t^n \; . 
\label{eq:gi19} 
\en 
Note that there are $4N$ links on the 4 dimensional lattice and that the 
factor $2^{4N} $ arises from the sum over $^\ast Z_{\ast l}$. 
Thus, using (\ref{eq:gi19}) in (\ref{eq:gi17}), we find 
\be 
{\cal Z} \; = \; \exp \left\{ 6N\,  \beta \right\} \; 2^{-4N} \;
\sum _{ \{ ^\ast Z_{\ast l} \} } \; \prod _{\ast p} 
\Bigl[ 1 \; + \; t \; P _{ \ast p} [^\ast Z ] \Bigr] \; , \hbo 
t \; = \; \exp \{ -2 \beta \} \; . 
\label{eq:gi20} 
\en 
Since the plaquette $P _{ \ast p} [^\ast Z ]$ only acquires values 
$\pm 1$, we may write: 
\be 
\prod _{\ast p} \exp \Bigl\{  \widetilde{\beta } \; 
P _{ \ast p} [^\ast Z ] \Bigr\} \; = \, 
[\cosh  \widetilde{\beta } ]^{6N} \; \prod _{\ast p} \; 
\Bigl[ 1 \; + \; \tanh  \widetilde{\beta } 
\; P _{ \ast p} [^\ast Z ] \Bigr] \; . 
\label{eq:gi21} 
\en 
The partition function (\ref{eq:gi20}) therefore becomes 
\bea 
{\cal Z} &=&  \left[ \frac{ \exp \{ \beta \} }{ 
\cosh  \widetilde{\beta } } \right]^{6N} \; 2^{-4N} \; 
\sum _{ \{ ^\ast Z_{\ast l} \} } \; 
\exp \Bigl\{  \widetilde{\beta } \; \sum _{\ast p} 
P _{ \ast p} [^\ast Z ] \Bigr\} \; , 
\label{eq:gi22} \\ 
\exp \{ -2 \beta \} &=&  \tanh  \widetilde{\beta } \; . 
\label{eq:gi23} 
\ena 
First of all we note that the dual of pure $Z_2$ gauge theory is another 
4-dimensional  $Z_2$ gauge theory: the model is self-dual. Furthermore, 
relation (\ref{eq:gi23}) is already familiar to us: 
we have obtained a complete analogue of the relation 
between $\beta $ and its dual $ \widetilde{\beta }$ for the 2d Ising model. 
We therefore once again encounter the fact that the weak coupling regime 
is mapped to the strong coupling regime of the same model. 
As a byproduct we find that the critical coupling is given by 
\be 
\beta _c \; = \; \frac{1}{2} \, \ln (1 + \sqrt{2}) \; \approx \; 
0.44068679 \ldots \; . 
\label{eq:gi24} 
\en 
The fact that the critical couplings of the 2d Ising model and 4d pure 
$Z_2$ gauge theory coincide might be a numerical accident. At least, 
I do not know any deeper reason for it. 
I finally point out that the $Z_2$ gauge symmetries of the original and the 
dual formulation are completely unrelated. This can be most easily seen 
from the fact that, at an intermediate stage, we have formulated 
the model entirely in terms of physical, i.e., gauge invariant 
variables: the closed vortex sheets of the dual lattice. Resolving 
this constraint, the gauge invariance of the dual model arose 
from a parameterisation invariance, namely, the redundancy 
when we performed the sum over all closed world sheets with the help 
of dual gauge links $^\ast Z$. 

\vskip 0.3cm 
{\tt 3 dimensions: } Let us finally discuss the 3-dimensional model. 
In 3 dimensions, a plaquette $p$ is mapped to a link $^\ast l$ and a cube $c$ 
is mapped to a site $^\ast x$ on the dual lattice. 
The constraint (\ref{eq:gi15}) then translates to 
\be 
\prod _{^\ast l \in ^\ast x} v_{^\ast l} \; = \; 1 \; . 
\label{eq:gi25} 
\en
The plaquettes carrying negative flux on the original lattice are 
represented by links forming closed loops on the dual lattice. 
The partition function now takes the form (there are $3N$ links on the 
lattice):
\be 
{\cal Z} \; = \; \exp \left\{ 3N\, \beta \right\} \; \sum _{ \{\hbox{closed 
loops} \} } \;  \exp \left\{ -2  \beta \, n\right\} \; . 
\label{eq:gi26} 
\en 
We already know how to perform the sum over all closed loops 
from subsection~\ref{sec:dual}: 
\bea 
&2^{-3N}& 
 \sum _{ \{ \tau_{\ast x} \} } \prod _{\ast l}  \; 
\exp \Bigl\{  \widetilde{\beta } \, \tau _{\ast x} \tau_{\ast y} \Bigr\} 
\label{eq:gi27}  \\
&=&  \Bigl[ \frac{ \cosh \widetilde{\beta } }{2} \Bigr] ^{3N} \; 
\sum _{ \{ \tau_{\ast x} \} } \; \prod _{\ast l} 
\Bigl[ 1 \; + \; \tanh \widetilde{\beta } \;  \tau _{\ast x} 
\tau_{\ast y} \Bigr] \; = \; 
\Bigl[ \cosh \widetilde{\beta } \Bigr] ^{3N} \; 
\sum _{ \{ \hbox{closed loops} \} } \; 
[ \tanh \widetilde{\beta } ]^n \; . 
\nonumber 
\ena 
Identifying once again 
\be 
\exp \{ -2 \beta \} \; = \;   \tanh  \widetilde{\beta } \; , 
\label{eq:gi28} 
\en 
we find 
\be 
{\cal Z} \; = \;  \left[  \frac{ \exp \{  \beta \} }{ 
2 \; \cosh \widetilde{\beta } } \right]^{3N} \; 
\sum _{ \{ \tau_{\ast x} \} } \; \exp \Bigl\{  \sum _{\ast l} \; 
\widetilde{\beta } \, \tau _{\ast x} \tau_{\ast y} \Bigr\} \; . 
\label{eq:gi29} 
\en 
Obviously, the $Z_2$ gauge theory is dual to a theory which is not a 
gauge theory anymore, the 3d Ising model. 
This has tremendous consequences: for the standard Ising model, 
cluster update algorithms are available. Using the Swendsen-Wang or 
Wolff type cluster update, we are able to simulate a gauge theory 
with much less autocorrelations. 
Unfortunately, such a framework is not (yet) available for 
more relevant theories such as lattice Yang-Mills theories.  

\subsection{Setting up lattice Yang-Mills theory } 

\begin{figure}[t]
\centerline{ 
\epsfxsize=10cm
\epsffile{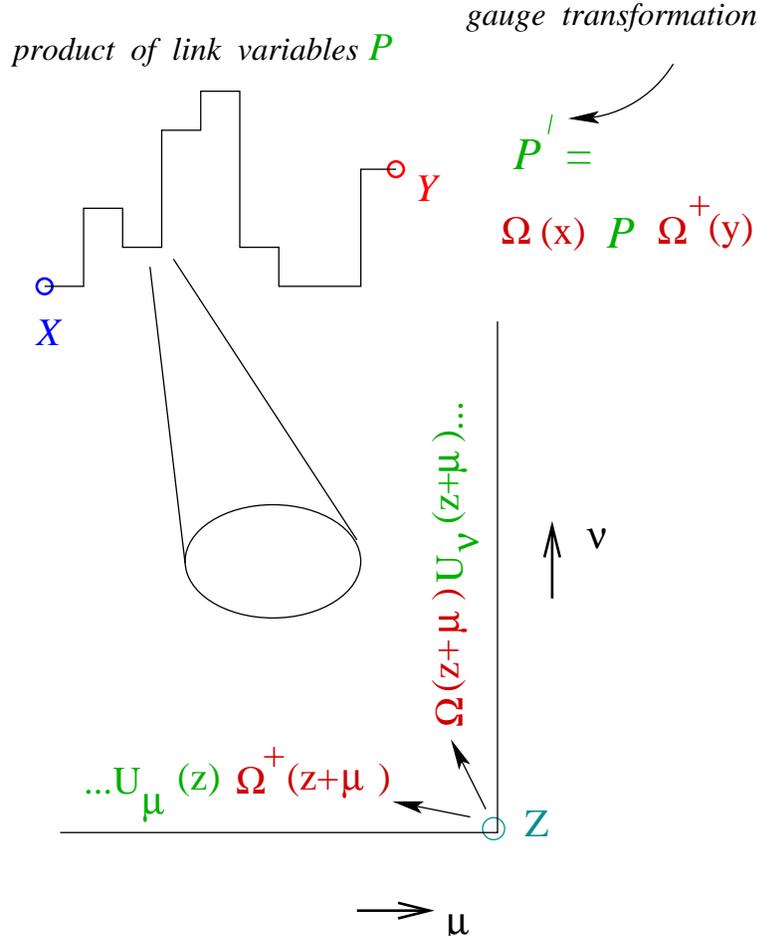}
}
\caption{ Path ordered product of link variables. } 
\label{fig:6} 
\end{figure}
Due to the universality conjecture, the lattice 
model with the correct number of dimensions and the correct 
symmetries uniquely defines the corresponding quantum field 
theory in the critical limit. The purpose of the present subsection 
is to propose a classical lattice model which satisfies these 
prerequisites in the case of Yang-Mills theory. 

\vskip 0.3cm 
The QCD matter fields (quarks) 
belong to the fundamental representation of the so-called $SU(N_c)$
colour group ($N_c=3$ for QCD). 
{\tt Gauge invariance} means that the action of the quark fields 
is invariant under the {\it local} unitary transformations, i.e., 
\be 
q(x) \rightarrow q^\prime (x) \; = \; \Omega (x) \, q(x) \, , \hbo 
\Omega (x) \in SU(N_c). \; . 
\label{eq:70} 
\en 
As explained in many text books, local gauge invariance of the quark 
kinetic term may only be achieved by introducing additional dynamical 
fields, the gluon fields $A_\mu (x) $. 

\begin{figure}[t]
\centerline{ 
\epsfxsize=10cm
\epsffile{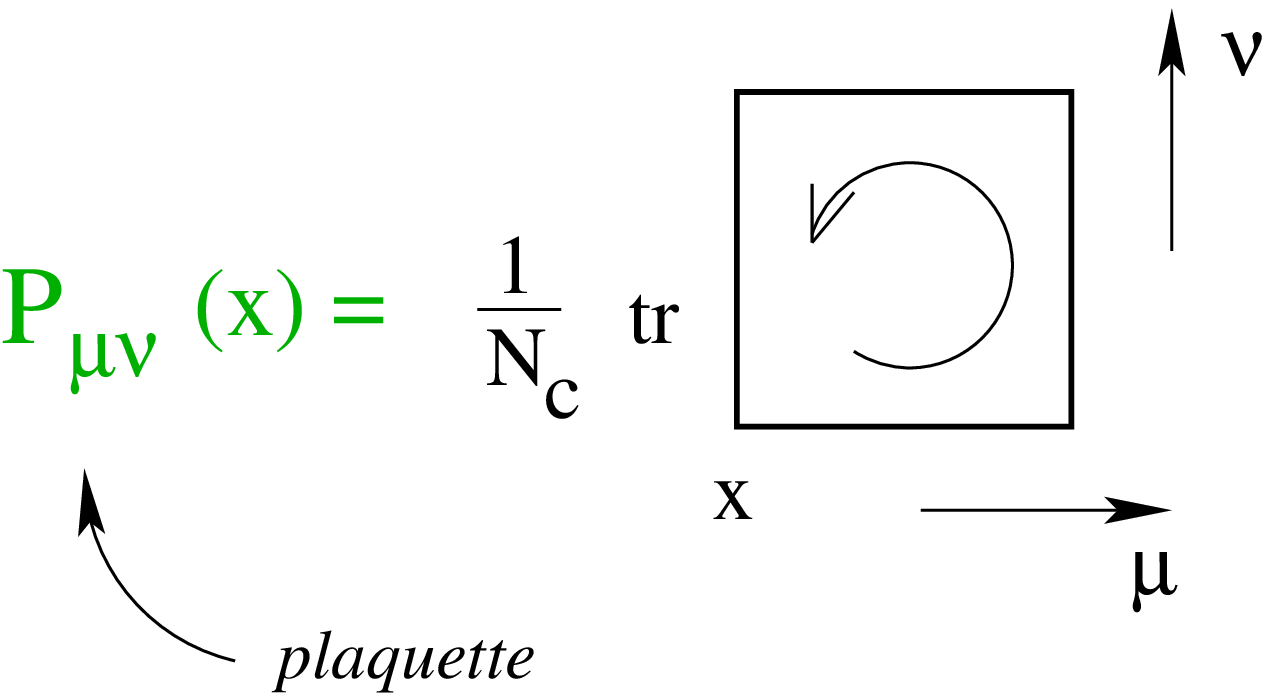}
}
\caption{ Lattice plaquette variable }
\label{fig:7} 
\end{figure}
\vskip 0.3cm 
The quark fields are associated 
with the sites in a lattice formulation. Hence, the symmetry group 
of the classical lattice Yang-Mills model is $[SU(N_c)]^{N_s}$, where 
$N_s$ is the number of lattice sites. In order to enforce such a high 
symmetry in the critical limit of a lattice model, it has turned out 
essential to realise the symmetry even for finite values of the 
lattice spacing $a$. This in turn forces the model to maintain local gauge 
invariance in the continuum limit~\cite{Wilson:1974sk}. A potential candidate 
for a quark kinetic term in the non-interacting case is 
\be 
\sum _{x, \mu } \frac{1}{2} \biggl[ 
\bar{q}(x) \, \gamma _\mu \, q(x+\mu ) \; - 
\bar{q}(x+\mu ) \, \gamma _\mu \, q(x ) \biggr] \; ,
\label{eq:71}
\en 
where the $\gamma _\mu $ are the Euclidean $\gamma $ matrices. 
Of course, the action (\ref{eq:71}) is not 
invariant under the gauge transformations (\ref{eq:70}). 
To achieve this invariance, we introduce an additional field 
of vector type, thus being related to the links 
of the lattice, 
\be 
U_\mu (x) \; \in \; SU(N_c) \; . 
\label{eq:72} 
\en 
Generalising the quark kinetic term (\ref{eq:71}) to 
\be 
S_Q \; = \; \sum _{x, \mu } \frac{1}{2} \biggl[ 
\bar{q}(x) \, \gamma _\mu \, U_\mu(x) \, q(x+\mu ) \; - 
\bar{q}(x+\mu ) \, \gamma _\mu \, U^\dagger _\mu(x) \, q(x ) \biggr]  \; , 
\label{eq:73}
\en 
one obtains the desired local invariance upon demanding that the link 
fields transform as 
\be 
U_\mu(x) \; \rightarrow \; \Omega (x) \, U_\mu(x) \, \Omega ^\dagger
(x+\mu) \; . 
\label{eq:74} 
\en 
Let us follow the case of the gauged Ising model and construct 
a kinetic term for the link fields $U_\mu(x)$. 
For lattice models ``kinetic'' means that the 
interactions of the fields on the lattice are {\it short range}, i.e., 
only nearest neighbours are involved. In order to design a gauge invariant 
kinetic term for every value of the 
lattice spacing, we firstly investigate the transformation properties 
of a path ordered product of link variables. 
Let us consider an open path $C$ 
which starts at point $x$ and ends at $y$ (see figure \ref{fig:6} 
for an illustration), and define
\be 
P(x,y) \; = \; { \cal P} \; \prod _{x\in C} \, U(x) \; , 
\label{eq:75}
\en 
where ${ \cal P} $ implies path ordering. 
Inserting the gauge transformed links (\ref{eq:74}) into
(\ref{eq:75}), one finds 
\be 
P(x,y) \; \rightarrow \; P^\prime (x,y) \; = \; \Omega (x) \; 
P(x,y) \; \Omega (y) \; . 
\label{eq:76} 
\en 
With the help of (\ref{eq:75}), it is easy to construct a kinetic 
term for the link variables which (i) is gauge invariant and 
(ii) involves only next to nearest neighbours. For this purpose 
one chooses $C$ to be a closed path starting at $x$ and ending at 
$y=x$ which encircles an elementary {\tt plaquette} (see figure \ref{fig:7}): 
\bea 
P_{\mu \nu }(x) &=& \frac{1}{N_c} \; \tr \, P(x,y) 
\nonumber \\ 
&=& \frac{1}{N_c} 
\tr \, \Bigl\{ U_\mu (x) \; U_\nu (x+\mu) \; U^\dagger _\mu (x+\nu) \; 
U^\dagger _\nu (x) \Bigr\} \; . 
\label{eq:77} 
\ena
Using (\ref{eq:76}) and the invariance of the trace under cyclic 
permutations, one easily shows that the plaquette 
(\ref{eq:77}) is indeed gauge invariant. 

\vskip 0.3cm 
The lattice partition function involves an integration over the 
dynamical fields of the theory. In the case of the group valued 
link variables (\ref{eq:72}), the question arises which measure 
${\cal D} U_\mu $ should be employed for the 
integrations. We must demand that the integration measure does not spoil 
gauge invariance. To ensure this we use the so-called Haar measure 
which satisfies 
\be 
d U_\mu (x) \; = \; d \biggl( A  U_\mu (x) B \biggr) \; , \hbo 
A, \, B \, \in \, SU(N_c) \; . 
\label{eq:78} 
\en 
The Haar measure is available in closed form for the unitary groups 
$SU(N_c)$. Here, I will only present the Haar measure for SU(2) 
group integrations. The SU(2) unitary matrix $U$ is conveniently 
parameterised in terms of the Pauli matrices, 
\be 
U \; = \; a_0 \; + \; i \, \vec{a} \vec{\tau } \; , \hbo 
U U^\dagger =1 \; \; \rightarrow \; a_0^2 + \vec{a}^2 \; = \; 1 \; . 
\label{eq:79} 
\en 
Since the constraint $U U^\dagger =1$, i.e. $a_0^2 + \vec{a}^2 =1$, is 
not changed if $U$ is multiplied with $A$ from the left and $B$ from
the right, respectively, these multiplications can be viewed as 
rotations in the 4-dimensional space spanned by $(a_0, \vec{a})$. 
Therefore, an invariant measure can be defined by 
\be 
dU \; = \; da_0 \; da_1 \; da_2 \; da_3 \; \delta\biggl( 
a_0^2 + \vec{a}^2 - 1\biggr) \; . 
\label{eq:80}
\en 
Introducing polar coordinates for the 3-dimensional vector 
$\vec{a} := a \vec{n} $, $\vec{n} \vec{n} =1$, the integration over 
the norm of the vector $\vec{a}$ 
can be performed with the help of the $\delta $ 
function in (\ref{eq:80}). We obtain the final result for the 
SU(2) Haar measure, i.e., 
\be 
dU \; = \; da_0 \; \sqrt{1-a_0^2} \; d \Omega _{\vec{n}} \; , 
\label{eq:81}
\en 
which is commonly used in lattice simulations.

\vskip 0.3cm 
Finally, the lattice representation of the gauge 
invariant partition function is given by
\be 
Z \; = \; \int {\cal D} U \; {\cal D} q \; {\cal D} q^\dagger  \; 
\exp \biggl\{ - S_Q \, + \, \beta \, \sum _{x, \mu > \nu } 
\frac{1}{2} \Bigl[ P_{\mu \nu }(x) + \mathrm{h.c.} \Bigr] \biggr\} \;
, 
\label{eq:82}
\en
where the quark interaction is encoded in $S_Q$ (\ref{eq:73}) 
and  $ P_{\mu \nu }(x)$ is the plaquette (\ref{eq:77}). 
$\beta $ is related 
to the bare gauge coupling constant $g$ of the continuum formulation 
by $\beta = 2 N_c/g^2$. 
The particular choice (\ref{eq:82}) of the 
lattice regularised gluonic action is known as the 
{\it Wilson action}~\cite{Wilson:1974sk}. 
Note that the fields $q(x)$, $q^\dagger (x)$ are anti-commuting 
Grassmann fields. This choice for the fermionic fields is necessary 
to obtain the correct Fermi statistics as well as to ensure 
the Pauli principle. It implies that the lattice model (\ref{eq:82}) 
cannot be straightforwardly be used in numerical simulations. 
Rather, since the action for the quark fields is quadratic, the
integration over the quark fields has to be performed analytically: 
\be 
\int  {\cal D} q \; {\cal D} q^\dagger  \; 
\exp \Bigl\{ - \bar{q} _A M_{AB} q_B \Bigr\} \; = \; \Det M[U] \; . 
\label{eq:83}
\en 
where the index $A$ comprises space-time as well as spinorial,
etc.~indices. The quark determinant $\Det M[U]$ is a gauge invariant 
function of the link variables $U_\mu (x)$. 
Note, however, that the link interaction mediated by the quark determinant 
is non-local, implying that a link at a particular site is coupled 
to all other links of the lattice. In practice, this implies that 
a local update of a single link enforces the calculation of a functional 
determinant. This explains why the numerical simulation of Yang-Mills 
theory with dynamical quarks requires much more computational resources 
than the simulation of the theory in {\tt quenched approximation}, 
where the quark determinant is neglected for the update of the 
link variables. 

\subsection{The fermion doubling problem } 

It turns out that the treatment of the quark degrees of freedom 
in (\ref{eq:82}) is still too naive: since the Dirac equation 
is linear in momentum, its lattice analogue does not produce just 
the desired quark degrees of freedom in the limit 
$a\rightarrow 0$, but rather $2^D$ copies of them ($D$ is the 
number of space time dimensions). This is already true 
for the free theory as will be shown in what follows. 

\vskip 0.3cm 
Let us firstly introduce the generating functional for connected 
Green functions in the case of free and massless bosonic theory, 
\be 
Z[j] \; = \; \int {\cal D} \phi \; \exp \biggl\{ - \frac{1}{2} 
\phi _k \Pi_{kl} \phi _l \; + \; j_x \phi _x \; \biggl\} \; . 
\label{eq:200} 
\en 
A sum is understood over indices which appear twice. One easily
verifies that the connected correlation function is obtained 
from $Z[j]$ via 
\be 
f(x-z) := \langle \phi _x \phi _z \Bigr\rangle \; - \; 
\Bigl\langle \phi _x \Bigr\rangle \Bigl\langle \phi _z \Bigr\rangle 
\; = \; \frac{ d \, \ln Z[j] }{ dj_x \; dj_z } \; . 
\label{eq:201} 
\en 
By ``completing the square'' in (\ref{eq:200}), we find 
\be 
Z[j] \; \propto \; \exp \biggl\{  \frac{1}{2} j_x \Bigl( \Pi ^{-1} 
\Bigr)_{xz} j_z \biggl\} \; , 
\label{eq:202} 
\en 
and hence for the free bosonic case 
\be 
\Bigl\langle \phi _x \phi _z \Bigr\rangle \; - \; 
\Bigl\langle \phi _x \Bigr\rangle \Bigl\langle \phi _z \Bigr\rangle 
\; = \; \biggl( \Pi ^{-1} \biggr) _{xz} \; . 
\label{eq:203} 
\en 
In order to evaluate the inverse $\Pi ^{-1}$, of the ''kinetic'' operator, 
we introduce its eigenvalues and eigenvectors, whereupon
\be 
\Pi \, \vert k \rangle \; = \; \lambda _k \, \vert k \rangle \; , 
\label{eq:204} 
\en  
and formally write 
\be 
\Bigl( \Pi ^{-1} 
\Bigr)_{xz} \; = \; \sum _k \, \vert k \rangle \; \frac{1}{\lambda _k} \; 
\langle k \vert \; . 
\label{eq:205} 
\en 
It is now easy to calculate the correlation function for the 
continuum case $\Pi = - \partial ^2$. The eigenfunctions are subjected 
to periodic boundary conditions $\phi (x) = \phi(x+L)$, i.e., 
\be 
\phi (x) \propto e^{ i kx } \;, \; , \hbo 
e^{ikL} = 1 \; , \hbo k = \frac{2 \pi }{L} n \; , \; \; n \in Z \; .
\label{eq:206} 
\en 
\begin{figure}[t]
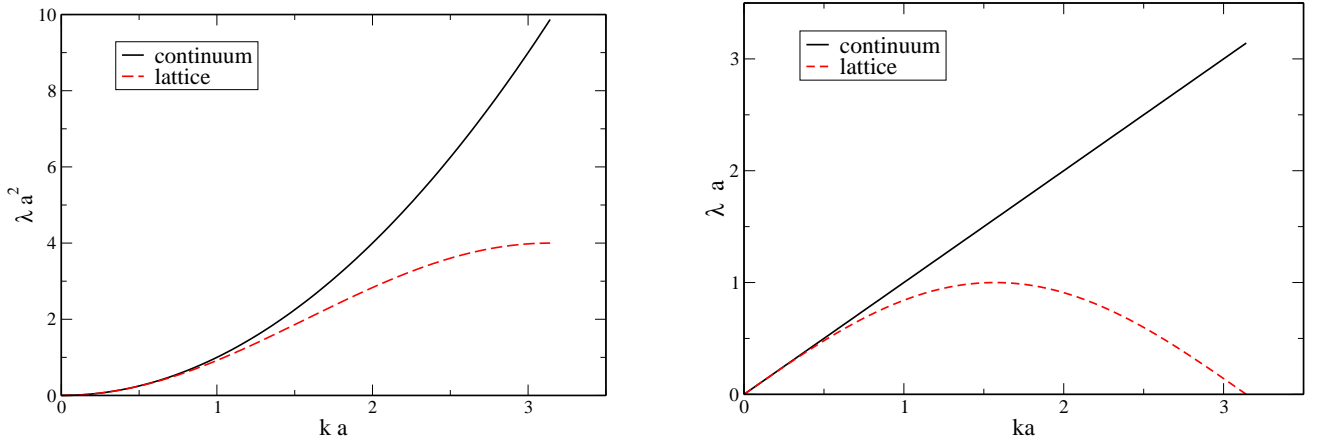

\centerline{ 
\epsfxsize=8cm
\epsffile{disper_b.eps} \hspace{1cm}
\epsfxsize=8cm
\epsffile{disper_f.eps}
}
\caption{ Dispersion relation from the tree level kinetic term (continuum
versus lattice formulation) for the bosonic case (left) 
and the fermionic case (right panel). }
\label{fig:7b} 
\end{figure}
The discrete $k$ levels are called {\tt Matsubara frequencies}. 
In the continuum, there are no further restrictions on the integer $n$. 
Making the ansatz (\ref{eq:206}), we find that the eigenvalues 
are given by 
\be 
\lambda (k) \; = \; k^2 \; \, \hbo \mathrm{(continuum)} \; . 
\label{eq:207} 
\en 
Hence, a free massless particle manifests itself in the correlation
function (\ref{eq:205}) as a pole at zero momentum transfer. 
The lattice version of the eigenvalue equation is 
\be 
\Pi \phi (x) \; = \; \sum _\mu \biggl[ - \, \phi (x+\mu ) \, 
+ \, 2 \, \phi(x) \, - \,  \phi (x-\mu ) \biggr] \; = \; 
\lambda _{latt} a^2 \; \phi (x) \; . 
\label{eq:208} 
\en 
In order to solve this equation, we use the plane wave ansatz
(\ref{eq:206}). One crucial difference between the lattice and 
the continuum version is that only wavelengths $l$ obeying 
\be
\frac{l}{2} \; \ge \; a \; , \hbo \frac{\pi}{k} \ge a 
\label{eq:209} 
\en 
are sensible. The lattice naturally provides an UV momentum cutoff,
i.e., $\Lambda _{UV} = \pi /a $. Inserting (\ref{eq:206}) into 
(\ref{eq:208}) one finds 
\be 
\lambda _{latt} a^2 \; = \; \sum _\mu \Bigl[ 2 \, - \, e^{ik_\mu a} 
\, - \, e^{-ik_\mu a} \Bigr] \; = \; 4 \, \sum_\mu \sin ^2 \Bigl( \frac{ 
k_\mu a }{2} \Bigr) \; . 
\label{eq:210} 
\en
For momenta which are small compared to the UV cutoff, i.e., $ka \ll
\pi $, we recover the continuum dispersion relation 
\be 
\lambda _{latt} \; = \; k^2 \, \Bigl[1 \, + \, {\cal O} (k^2a^2) 
\Bigr] \; . 
\label{eq:211} 
\en 
In figure \ref{fig:7b} the continuum dispersion relation 
for bosons is compared to its lattice version. 
The lattice  correlation function has only 
one singularity reflecting that in the scaling limit $\lambda a^2 
\ll 1$, $ka \ll \pi $, the dispersion relation of the continuum free particle 
is recovered. 

\vskip 0.3cm 
Let us move on to the fermionic case. In order to reproduce the correct 
Fermi statistics, fermion fields $\psi (x)$ are of Grassmann type and 
obey anti-periodic boundary conditions. I refer to the
textbook~\cite{bell}
for an introduction to the free fermionic theory, and only quote 
the final result for the correlation function which formally 
agrees with (\ref{eq:205}). In the continuum, the eigenvalue 
equation is given by 
\be 
\Pi \psi (x) \; = \; \dslash \psi (x) \; = \; \lambda \, \psi (x) 
\; , 
\label{eq:212} 
\en 
where anti-hermitian (Euclidean) $\gamma $ matrices are used. 
The ansatz for the spinor wave functions is again of plane wave type, 
\be 
\psi (x) \propto  u(k) \; e^{ i kx } \;,  \hbo 
e^{ikL} = -1 \; , \hbo k = \frac{2 \pi }{L} \Bigl( n + \frac{1}{2}
\Bigr) \; , \; \; n \in Z \; .
\label{eq:213} 
\en 
The spectrum $\lambda (k)$ is determined by making the ansatz 
\be 
u(k) \; = \; \Bigl[ i\kslash \: + \; \lambda \Bigr] \; u_0 \; , 
\en 
which yields 
\be 
\Bigl[ i \kslash - \lambda \Bigr] u(k) \; = \; 
\Bigl[ i \kslash - \lambda \Bigr] \Bigl[  i \kslash + \lambda \Bigr] 
u_0 \; = \; 0 \; , 
\en 
and therefore 
\be 
\Bigl[ k^2 - \lambda^2 \Bigr] u_0 \; = \; 0 \; . 
\en 
Hence, the spectrum of the continuum theory is linearly increasing, 
$\lambda = \pm \sqrt{k^2}$. 
Using the kinetic energy for a free quark theory introduced in 
(\ref{eq:71}), the lattice version of the eigenvalue equation is given
by 
\be 
\frac{1}{2} \sum _\mu \biggl[ \gamma _\mu \, \psi (x+\mu) \, - \, 
\gamma _\mu \, \psi (x - \mu ) \biggr] \; = \; \lambda \, a \; \psi (x) \; . 
\label{eq:214} 
\en 
The ansatz (\ref{eq:213}) also provides the eigenvectors of the 
eigenvalue problem (\ref{eq:214}). 
Repeating the steps which have led to the continuum dispersion
relation, one finds its lattice analogue
\be 
\lambda \,a \; = \; \sqrt{ \sum _\mu \sin ^2 \Bigl( k_{\mu} a \Bigr) } \;
. 
\label{eq:215} 
\en 
The fermionic eigenvalue distribution is shown in figure 
(\ref{fig:7b}), right panel. Close to the critical limit 
$(\lambda a \ll 1)$, one recovers the continuum dispersion relation from 
(\ref{eq:215}) by making a Taylor expansion with respect to $ka$. 
In addition, a second singularity occurs for $ka \approx \pi $. 
This shows that even if $\lambda a \ll 1$ a second 
fermion flavour arises from the lattice fermion action (\ref{eq:71}).

\vskip 0.3cm 
It can be shown that this 
fermion doubling problem must occur for a chirally invariant 
action which is translationally invariant and local ({\tt 
Nielsen-\-Ninomiya  No-Go theorem}). At the present stage, a lot of 
research effort is devoted to incorporate chiral symmetry at the
expense of, say, a moderate non-locality of the action~\cite{kaplan}.

\subsection{Overlap fermions} 

In the continuum formulation, the chirally invariant Dirac operator 
${\cal D}$ satisfies the relations 
\be 
\bigl\{ {\cal D}, \gamma _5 \bigr\} \; = \; 0 \; , \hbo 
\bigl\{ {\cal D}^{-1}, \gamma _5 \bigr\} \; = \; 0 \; , 
\label{eq:o1} 
\en 
which tells us that the non-zero eigenvalues $\bar{ \lambda }$ 
appear in pairs $\{ \bar{ \lambda }, - \bar{ \lambda } \}$. 
Let $D$ denote a lattice candidate of the Dirac operator (in units 
of the lattice spacing $a$) satisfying the 
so-called { \it Ginsparg-Wilson relation}~\cite{Ginsparg:1981bj},
\be 
\bigl\{ D , \gamma _5 \bigr\} \; = \; 2 \, D \; \gamma _5 \; D \; .
\label{eq:o2} 
\en 
One observes that the right hand side of (\ref{eq:o2}) is of order 
$a^2$ (compared with the order $a$ of the 
left hand side) implying that the naive continuum limit 
$a \rightarrow 0$ of (\ref{eq:o2}) reduces to the chiral relation 
(\ref{eq:o1}). The most important observation, however, is that a
certain remnant of the chiral symmetry is still present in the lattice 
version. Defining 
\be 
\widetilde{D}^{-1} \; := \; D^{-1} \; - \; 1 \; , 
\label{eq:o3} 
\en 
and using 
$$ 
\bigl\{ \gamma _5, D^{-1} \bigr\} \; = \; 2 \, \gamma _5 \; , 
$$ 
which directly follows from the Ginsparg-Wilson relation (\ref{eq:o2}),
we observe that 
$\widetilde{D}^{-1} $ may be used as a chirally invariant 
quark propagator, i.e., 
\be
\bigl\{ \gamma _5, \widetilde{D}^{-1} \bigr\} \; = \; 0 \; . 
\label{eq:o4} 
\en 
Hence, we are left with 
the task to find an operator $D$ obeying the Ginsparg-Wilson 
relation (\ref{eq:o2}). Here, I will briefly discuss the Overlap 
Dirac operator~\cite{Narayanan:1994gw}-\cite{Edwards:1998wx}, 
firstly introduced in the pioneering paper~\cite{Narayanan:1994gw} by 
Narayanan and Neuberger. One introduces 
\be 
D \; = \; \frac{1}{2} \Bigl[ 1 \; + \; \gamma _5 \, H \Bigr] \; , 
\label{eq:o5} 
\en 
where $H$ is a Hermitian operator with eigenvalues $\pm 1$. 
Common choice is 
\be 
H \; = \; D_w \; / \; \Bigl( D_w ^\dagger D_w \Bigr) ^{1/2} \; , 
\label{eq:o6} 
\en 
where $D_w$ is the standard Hermitian Wilson-Dirac operator. 
Inserting (\ref{eq:o5}) into (\ref{eq:o2}), it is 
straightforward to prove that $D$ from (\ref{eq:o5}) satisfies the 
Ginsparg-Wilson relation (\ref{eq:o2}). 
A comprehensive discussion of the quark propagator 
(\ref{eq:o3}) in the context of a simulation of SU(3) 
Yang-Mills theory can be found in~\cite{Bonnet:2002ih}.

\subsection{Measuring observables } 

\begin{figure}[t]
\centerline{ 
\epsfxsize=15cm
\epsffile{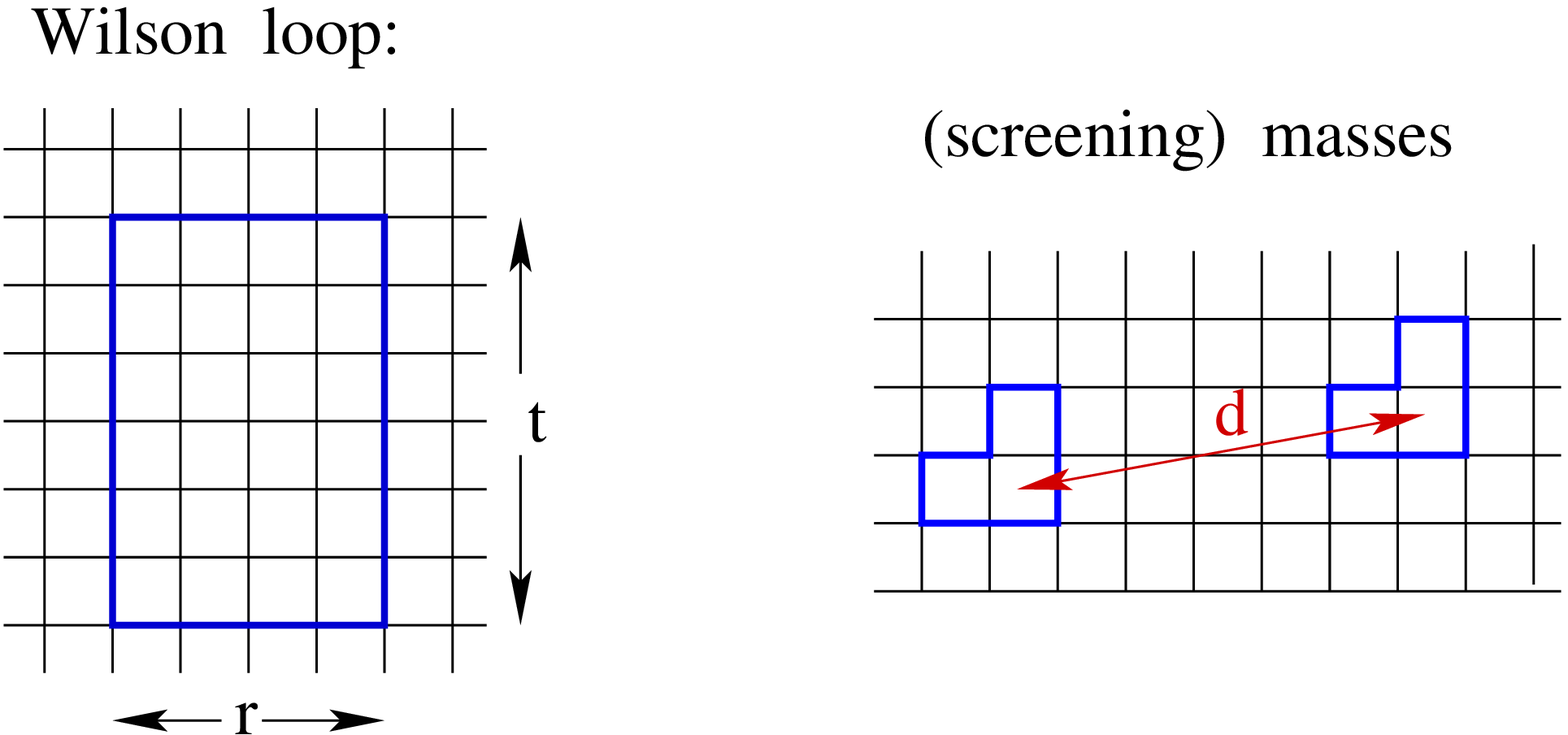}
}
\caption{ Wilson loop and loop--loop correlation function }
\label{fig:8} 
\end{figure}
We have observed that the trace of the path ordered product $P(x,y)$ of link 
variables (\ref{eq:75}) is gauge invariant when taken along a closed 
curve $C$, i.e., $x=y$. Depending on the choice for the loop 
$C$, the expectation value of such loop variables can be 
connected to physical observables. For instance, for the so-called
Wilson loop, we choose a rectangular loop with size $r$ in one spatial 
direction and the extension $t$ in the Euclidean time direction 
(see figure \ref{fig:8}, left panel). In the limit of large $t$, the 
Wilson loop expectation value is related to the potential $V(r)$
between a static quark and a static anti-quark which are separated by the 
distance $r$, i.e., 
\be 
\Bigl\langle W[C] \Bigr\rangle 
\; \propto \; \exp \Bigl\{ - V(r) \; t \Bigr\} \; , 
\label{eq:90} 
\en 
In the particular case that the potential is linearly rising, 
$V(r) = \sigma r $ with {\tt string tension} $\sigma $, 
one observes that the Wilson loop expectation value exponentially 
decreases with the area $A$ enclosed by the loop $C$. 
Since a linearly rising quark anti-quark potential implies 
confinement (see discussion below), this {\tt area law} (due to Wilson) 
is a litmus test for quark confinement. 

\vskip 0.3cm
Furthermore, one can calculate the correlation function $L(t_x-t_y, 
\vec{x}-\vec{y})$ 
of two loops centred at $x$ and $y$, respectively (see figure
\ref{fig:8}), right panel). Here, information is transported from point $x$ 
to $y$ by gauge invariant states $\vert ph \rangle$ . The 
shape of a particular loop 
determines its behaviour under the symmetry transformations of 
the underlying lattice. These symmetry transformations correspond 
to rotations in the continuum limit. Therefore, it is possible to 
select the spin quantum number of the state $\vert ph \rangle$ by 
adjusting the shape of the loop. For large distances $\Delta
= t_x-t_y$, the correlation function exponentially decreases, i.e., 
\be 
\sum _{\vec{u}} \, L(t_x-t_y, \vec{u}= \vec{x}-\vec{y}) \; \propto \; 
\exp \Bigl\{ - \, ma \; \Delta \Bigr\} \; . 
\label{eq:91}
\en 
Hence, the calculation of loop correlation functions provides access to 
the so-called {\tt screening masses} $m$ of physical particles. 
In the purely gluonic theory, the only gauge invariant states are the 
glue balls, while in full QCD also hadronic states contribute to 
the correlation functions. 

\subsection{The continuum limit } 

\begin{figure}[t]
\centerline{ 
\epsfxsize=10cm 
\epsffile{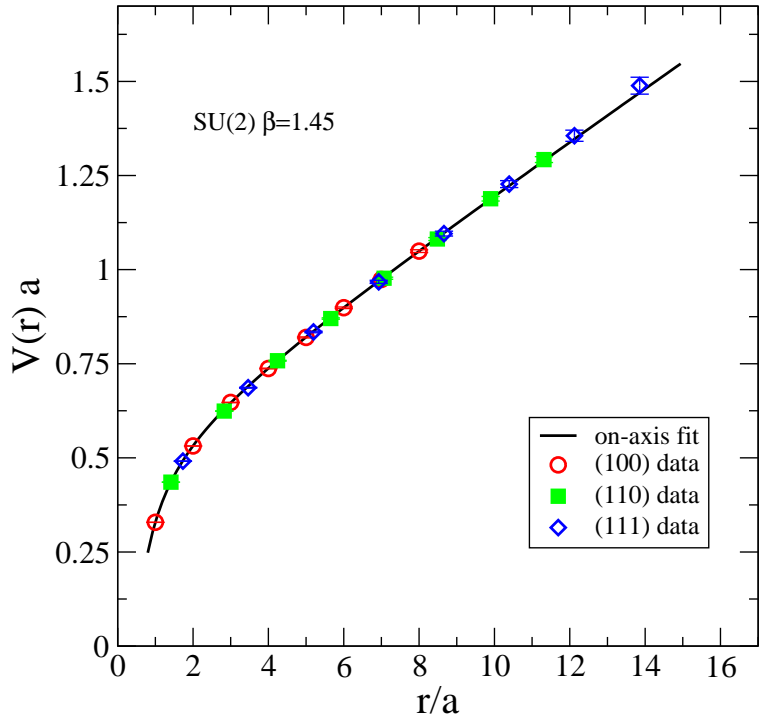}
}
\caption{ The static quark anti-quark potential as obtained from  pure 
SU(2) lattice gauge theory. Plot from~\cite{Langfeld:2007zw}. }
\label{fig:10} 
\end{figure}
For definiteness, I confine myself to the case of pure 
(i.e.~no quarks) SU(2) gauge theory. The generalisation of 
the findings of the present section to $SU(N_c)$ is straightforward. 
The task is now to find the critical 
limit of the lattice Yang-Mills theory. 

\vskip 0.3cm
There is a lesson to learn from continuum Yang-Mills theory. In order 
to renormalise the continuum theory, one absorbs a logarithmic 
divergence into the bare gauge coupling. A detailed calculation 
yields 
\be 
\frac{1}{g^2(\Lambda ) } \; = \; \frac{11}{24 \pi ^2 } \; 
\ln \frac{\Lambda ^2 }{\mu ^2} \; + \; \mathrm{finite} \; , 
\label{eq:92} 
\en 
where $\Lambda $ is the UV cutoff and where $\mu $ is an arbitrary 
renormalisation point. The coefficient in front of the logarithmic
term is the quantity of interest and can be obtained by evaluating a 
bunch of one-loop Feynman diagrams. Eq.(\ref{eq:92}) shows that in 
the critical limit $\Lambda \rightarrow \infty$ the bare coupling 
vanishes. This is one manifestation 
of the celebrated property of {\tt asymptotic 
freedom}. Switching from the continuum to the lattice formulation we 
identify $\Lambda = \pi /a $ and use $\beta =4/g^2$ to straightforwardly 
derive 
\be
a^2 (\beta ) \; = \; \mathrm{const.} \; \exp \biggl\{ 
-\frac{6 \pi ^2 }{11} \beta \biggr\} \; . 
\label{eq:93} 
\en 
Due to asymptotic freedom, we expect that the critical limit is 
approached when $\beta \rightarrow \infty $. The perturbative 
relation between $a$ and $\beta $ in (\ref{eq:93}) is called 
{\it asymptotic scaling}. 

\vskip 0.3cm 
Modern computer simulations use a more complicated ``kinetic'' term 
for the gluon fields. One example of such an {\tt improved action} 
is given by 
\be 
S \; = \; \beta \; \sum _{\mu > \nu, x} \Bigl[ \kappa _1 \; 
\bar{P}_{\mu \nu }(x) \; + \;  \kappa _2 \; \bar{P}^{(2)}_{\mu \nu }(x) 
\Bigr] \; . 
\label{eq:ai8} 
\en
where $ \bar{P}^{(2)}_{\mu \nu }(x) $ is the $2\times 2$ Wilson loop. 
Imposing the constraint 
\be 
\kappa _1 \; + \; 16 \; \kappa _2 \; = \;  1 \; , 
\label{eq:ai9} 
\en
ensures that the familiar relation between $\beta $ and the bare 
gauge coupling $g$, $\beta = 2N_c/g^2$, is maintained. 
The residual freedom of choosing $ \kappa _1 $ and  $ \kappa _2 $ 
can be used to obtain a rather good agreement with 
asymptotic scaling on rather coarse lattices. 

\vskip 0.3cm 
\begin{figure}[t]
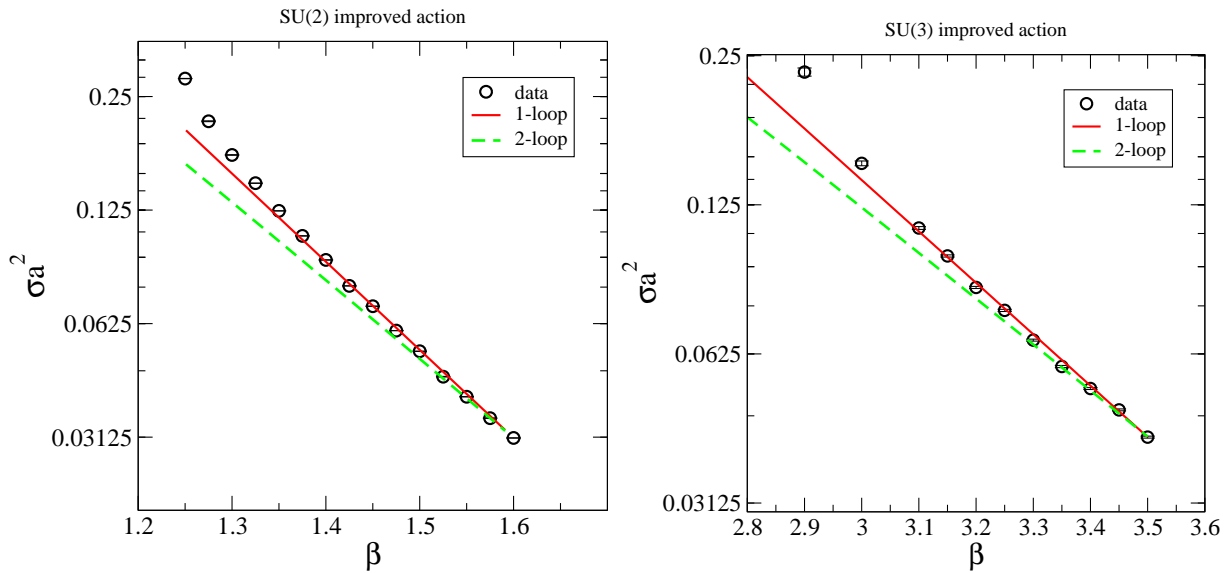

\centerline{ 
\epsfxsize=8cm
\epsffile{scaling_imp_su2_log.eps}
\epsfxsize=8cm
\epsffile{scaling_imp_su3_log.eps}
}
\caption{ Approaching the  continuum limit of SU(2) (left) and 
SU(3) (right) lattice gauge theory 
(improved action  from~\cite{Langfeld:2007zw}).
 } 
\label{fig:9} 
\end{figure}
In order to search for the critical limit 
with the help of numerical simulations, we calculate a physical 
quantity, e.g.~the string tension $\sigma $ in units of the lattice 
spacing as a function of the only parameter $\beta $. 
This is done by calculating the static quark anti-quark potential 
$V(r)$ as a function of the quark anti-quark distance $r=n \, a$. 
The outcome in units of the lattice spacing is shown in 
figure~\ref{fig:10}. By fitting the numerical data to 
$$ 
V(r) a \; = \; v_0 \; - \; \frac{\alpha }{n} \; + \, 
\sigma a^2 \; n \; , 
$$
we find the string tension in units of the lattice spacing, 
$\sigma a^2$, for each value of $\beta $. 
The outcome of this calculation  is shown in figure \ref{fig:9}. 
One indeed observes that the c-number $\sigma a^2$ exponentially
decreases for large values of $\beta $ in agreement with the 
prediction (\ref{eq:93}) of continuum Yang-Mills theory. 
The quantum field theoretical 
limit of the classical lattice model is obtained by interpreting 
the correlation length, i.e., the string tension $\sigma $ in the present 
example, as a fixed physical quantity, and reinterpreting the $\beta $ 
dependence of the numerical data for $\sigma a^2 $ as the $\beta $ 
dependence of the lattice spacing. 

\vskip 0.3cm 
Let us assume we have obtained a glue ball mass $m$ 
in lattice units, i.e., we know $ma$ as a function of $\beta $. 
If the mass $m$ is a physical observable, one must recover from the data 
the characteristic dependence $a(\beta )$ (see (\ref{eq:93})) for 
sufficiently large $\beta $ values. Hence, the ratio of the two 
dimensionless numbers $m^2 a^2 / \sigma a^2 $ approaches a constant 
for $\beta $ close to the critical point (see figure \ref{fig:9}, right 
panel). Extrapolating the data to the continuum limit $a \rightarrow 0$, 
i.e., $\beta \rightarrow \infty $, one determines the physical mass 
$m$ in units of another physical scale, i.e., $\sqrt{\sigma }$. 
Finally, let us count the number of parameters. The only parameter 
of the classical lattice model is $\beta $, but $\beta $ is no 
longer at our disposal in the quantum field theory limit (which 
implies $\beta \rightarrow \infty $). However, the physical value 
of the correlation length (or $\sqrt{\sigma }$ in the present 
example) takes over the role of a free parameter. The replacement of a 
dimensionless parameter by a mass scale in the continuum limit is 
a feature of many quantum field theories and is called 
{\tt dimensional transmutation}. On 
the lattice every mass scale is obtained in units of the string 
tension, $\sqrt{\sigma } = 440 \, $MeV is used to assign the familiar 
units of QCD to observables. 
For $32$ lattice points in any space-time direction, we then find:

\begin{center}
\begin{tabular}{l|cccc} 
$\beta $  (input)       & $1.250$    & $1.400$     & $1.500$     & $1.600$   \\
\hline 
$\sigma a^2$ (calculated)     & $0.279(2)$ & $0.0922(7)$ & $0.0528(3)$ & $0.0311(2)$ \\
$L=Na$            & $7.7 \,$fm & $4.4 \,$fm  & $3.3 \,$fm  & $2.6 \,$fm  \\
$\Lambda = \pi/a$ & $2.6 \,$GeV& $4.6 \,$GeV & $6.0 \,$GeV & $7.8 \,$GeV 
\end{tabular} 
\end{center}
For a fixed number of lattice points, we note that we cannot make 
$\beta $ arbitrarily small since the physical volume becomes 
too small. Small values of $\beta $ result in large volumes, but 
we cannot make $\beta $ too small in order to have a reasonably large 
UV cutoff. Thus, for a fixed number of points, there is a small 
window of $\beta $ values which are appropriate for a study of 
QCD particle properties. This window is sometimes called the {\it scaling 
window}.

\bigskip 

{\bf Acknowledgements: } I thank Tom Heinzl and Martin Lavelle 
  for a careful reading of the manuscript and helpful comments.

\end{document}